**A Critical Assessment of Electronic Structure Descriptors for Predicting Perovskite Catalytic Properties**

**Authors:** Ryan Jacobs[1,*], Jian Liu[2], Harry Abernathy[2], Dane Morgan[1]

[1] Department of Materials Science and Engineering, University of Wisconsin-Madison, Madison, WI, 53706, USA.

[2] National Energy Technology Lab, Morgantown, WV, 26505, USA.

*Corresponding author e-mail: rjacobs3@wisc.edu

## Abstract

The discovery and design of new materials which can efficiently catalyze the oxygen reduction and evolution reactions at reduced temperatures is important for facilitating the widespread adoption of fuel cell and electrolyzer technologies. Numerous studies have produced correlations between catalytic properties, such as oxygen surface exchange or electrode area specific resistance (ASR), and properties of the catalyst material. However, correlations have historically been limited in scope (e.g., using only a few materials or at a single temperature) and it has been difficult to provide detailed assessments of their robustness. Here, we assess the ability of the O p-band center electronic structure descriptor, obtained from density functional theory (DFT) calculations, to correlate with oxygen surface exchange rates, diffusivities, and area specific resistances for a large database of perovskite oxide catalytic properties. By data mining the literature, we obtain 747 catalytic property value data points spanning 299 unique perovskite compositions from 313 studies. We assess linear correlations of each property with the O p-band center and find generally modest correlations that are qualitatively useful (prediction mean absolute errors of about 0.5 log units are typical), where the correlations are improved at higher temperatures (e.g., 800 °C vs. 500 °C) and significantly improve when considering fits to the subset of materials which have multiple independent measurements. These findings suggest that the spread of property data is significantly influenced by experimental uncertainty, and subsequent measurements of additional materials will likely improve the O p-band center correlations.

# 1. Introduction

Oxygen-active materials that efficiently reduce, absorb, transport, and evolve oxygen are key components for energy devices such as solid oxide fuel cells (SOFCs), solid oxide electrolyzers (SOECs), reversible SOFCs (r-SOFCs), proton ceramic fuel cells (PCFCs), [1–11] solid oxide batteries,[12–14] memristors,[15–17] and oxygen separation membranes for high purity gas generation.[18,19] In this work, we focus on the perovskite oxide family of oxygen-active materials ($A_{1-x}A'_xB_{1-y}B'_yO_3$ chemical formula). Perovskite oxides are the most explored candidate materials to replace precious metal catalysts for use in SOFCs and SOECs to catalyze the oxygen reduction reaction (ORR) and oxygen evolution reaction (OER), respectively. A tremendous body of research has been conducted in the past few decades, resulting in novel perovskite compounds for application as the electrode in a fuel cell and electrolyzer devices as well as the oxygen-transport mediating electrolyte material. Regarding the electrolyte, doped zirconia and ceria materials are often employed (e.g. yttria-stabilized zirconia (YSZ) as well as gadolinium- or samarium-doped ceria (GDC, SDC)), but occasionally Sr- and Mg-doped $LaGaO_3$ perovskite, $La_{1-x}Sr_xGa_{1-y}Mg_yO_3$ (LSGM) is used, which remains one of the best oxygen-conducting electrolyte materials.[20–22] Perovskites have found more widespread use as the electrode material, where research has resulted in the development of current commercial SOFC materials like $La_{1-x}Sr_xMnO_3$ (LSM)[23–25] and $La_{1-x}Sr_xCo_{1-y}Fe_yO_3$ (LSCF),[26–28] and high-performing but presently non-commercial materials like $Ba_{1-x}Sr_xCo_{1-y}Fe_yO_3$ (BSCF)[29–33] and $PrBaCo_2O_{5+\delta}$ (PBCO).[34–36] Recently, it has been shown that some perovskites, like $BaFe_{0.4}Co_{0.4}Zr_{0.1}Y_{0.1}O_3$ (BFCZY),[5,37,38] $Ba_{0.95}La_{0.05}Fe_{0.8}Zn_{0.2}O_3$ (BLFZn)[10,39] and $SrFe_{1.5}Mo_{0.5}O_{6-\delta}$,[40,41] are highly oxygen-active materials which are also good incorporators and conductors of protons. Due to their ability to also transport electrons efficiently, these materials form a group of so-called triple conductors. Triple conducting materials are especially promising for use as electrodes in r-SOFCs and PCFCs, where the electrode must efficiently reduce and transport oxygen, protons and electrons simultaneously.[42–45]

In the past decade, a number of atomistic-scale computational investigations have been conducted to understand perovskite catalytic properties, primarily using first-principles density functional theory (DFT) methods. These computational studies sought to understand the origins of, and trends in, perovskite properties relevant to their desirable oxygen-active properties. In such studies, electronic structure descriptors are commonly used to formulate and rationalize correlations between material structure and chemistry and a corresponding property of interest, like the oxygen surface exchange rate. Electronic structure descriptors are useful in part because they serve as a simple proxy for a more complex, difficult-to-acquire physical property, and they are important because they have enabled new understanding and a means to screen and predict new materials using DFT. While many descriptors have been proposed in the literature, such as O p-band center,[46] O p-band center weighted by the $d_{xy}$ orbital band center,[47] $e_g$ electron orbital occupancy,[48,49] charge transfer gap energy,[50] and $G_{max}$, the maximum free energy of OER steps at zero overpotential,[51] among others, here we focus on the O p-band center as it has been previously shown to correlate well with a number of oxygen-active materials properties.[52,53] Properties shown to correlate with O p-band center include experimental surface exchange rates in perovskites and Ruddlesden-Popper oxides,[46,54,55] oxygen vacancy formation and migration energies,[46,52,56] calculated and experimental barriers for oxygen diffusion and surface exchange,[46,53,55] the electronic work function,[57,58] OER current densities and overpotential,[47,53,59] and the surface binding energies of numerous species such as O, H, OH, OOH. [52,53,60] The O p-band center has not only shed light on the above-mentioned correlations with material composition, but has also resolved trends in the coupling of lattice strain with changes in the electronic structure and resulting properties,[61,62] as well as distinguished surface-specific properties of work function[58] and adsorption energies.[60] There is also evidence suggesting that the O p-band center idea may be extensible beyond oxides as a more general anion p-band center descriptor, where it may find application to predict catalytic properties of sulfides, chlorides, and fluorides.[52] Further, the O p-band center correlation has enabled high-throughput DFT screening studies, such as the work by Jacobs et al.[55] who found new stable, highly active SOFC cathode materials[63,64] and Ma et al.,[57] who discovered a number of low work function perovskite materials for electron emission applications. While high-throughput computational screening using the O p-band center is more efficient than large-scale experimentation, a single prediction still requires at least one DFT calculation, which typically takes on the order of hours of computing time.

While previous studies employing descriptors like the O p-band center have resulted in new understanding and trends which have produced new promising materials, studies which have examined correlations of the O p-band center with various properties tended to either focus on a limited set of the examined materials from the literature, provided values obtained under a single set of experimental conditions (e.g., one particular temperature), or lacked detailed estimates of the uncertainty both from the experimental values and predicted values from the correlations. These factors, in turn, raise questions about how accurate the O p-band descriptor is for calculating oxygen-active materials properties and whether such correlations are more suited for qualitative materials screening guidance rather than quantitative predictions of materials behavior. The first major result of this work is development and exploration of a large database of experimentally measured perovskite catalytic properties from the literature, including tracer and chemical surface exchange rates k* and $k_{chem}$, tracer and chemical oxygen diffusivities D* and $D_{chem}$, and the electrode cell area specific resistance (ASR). This database is expected to be widely useful for the broader research community for data-driven analysis, for example, assessing typical ranges of values of the most-studied materials, providing guidance for expected error levels of properties measured by different research groups on the same material, flagging potentially problematic measurements, forming lists of high-performing materials at various working temperatures, and more. The second main result of this work consists of using our new database to explore and quantitatively assess many previously proposed correlations, including the trends of k* and ASR with O p-band center, trend of k* versus D*, and the relationships between $D^*/D_{chem}$ and $k^*/k_{chem}$. We find that O p-band center trends worsen compared to previous studies with the addition of more data, but significantly improve when considering the subset of materials which have multiple experimental measurements. This finding suggests that experimental error plays a key role in determining the correlation quality, and that the O p-band center is still useful for qualitative assessment of perovskite catalytic properties.

# 2. Results and Discussion

## 2.1. Experimental property database development

Catalytic property data of perovskite oxide materials were collected from a total of 313 studies in the literature (the full database is available as part of the **Supplementary Information (SI)**, see **Data Availability**) (see **Methods** in **Section 4.2**). We believe this database of perovskite catalytic properties is the largest such database which has been collected to date. The data consist of chemical surface exchange ($k_{chem}$) and diffusion ($D_{chem}$) values from electrical conductivity relaxation (ECR) experiments, tracer surface exchange (k*) and diffusion (D*) values from isotope exchange depth profiling (IEDP) experiments, and ASR measurements from electrochemical impedance spectroscopy (EIS) experiments. From these 313 studies, a total of 749 data points on 299 unique perovskite compositions were obtained. For the properties of $k_{chem}$, $D_{chem}$, k*, D* and ASR, a total of 98, 83, 80, 66 and 422 experimental measurements on various materials were obtained, respectively, which comprise 62, 58, 48, 42 and 257 unique materials for each property, respectively, as summarized in **Table 1**. As a visual overview of the collected data, in **Figure 1** we show a periodic table heatmap summarizing the occurrences of each element in our database. From **Figure 1**, we can observe that La is the most occurring rare earth element, and Sr, followed by Ba, are the most occurring alkaline earth elements. The elements Co and Fe are the most frequently occurring transition metal elements, and Nb is the most frequently occurring non-3d transition metal stabilizing element. The high frequency of these listed elements makes sense given the history of engineering new mixed ionic-electronic conductors, where perovskite $LaCoO_3$ is a high-performing parent material which is often modified for increased activity by A-site doping with Sr or Ba and B-site doping with Fe and Nb. In **Figure 2** we show histograms of database catalytic property values at 500 °C and 800 °C, in conjunction with some basic statistics of the catalytic property distributions. Additional details and discussion of aspects of the collected database are presented in **Section S1** of the **SI**. This supplementary discussion includes an overall survey of basic statistics of each catalytic property in the database, including flagging potentially problematic measurements and enumerating the best-performing materials at various temperatures based on each catalytic property (e.g., highest $k_{chem}$ materials, lowest ASR materials).

**Table 1.** High-level summary of perovskite catalytic data extracted from the literature for this study.

| Property | Number of studies examined | Number of measurements extracted | Number of unique materials |
|---|---|---|---|
| $k_{chem}$ | 70 | 98 | 62 |
| $D_{chem}$ | 56 | 83 | 58 |
| k* | 39 | 80 | 48 |
| D* | 37 | 66 | 42 |
| ASR | 235 | 422 | 257 |

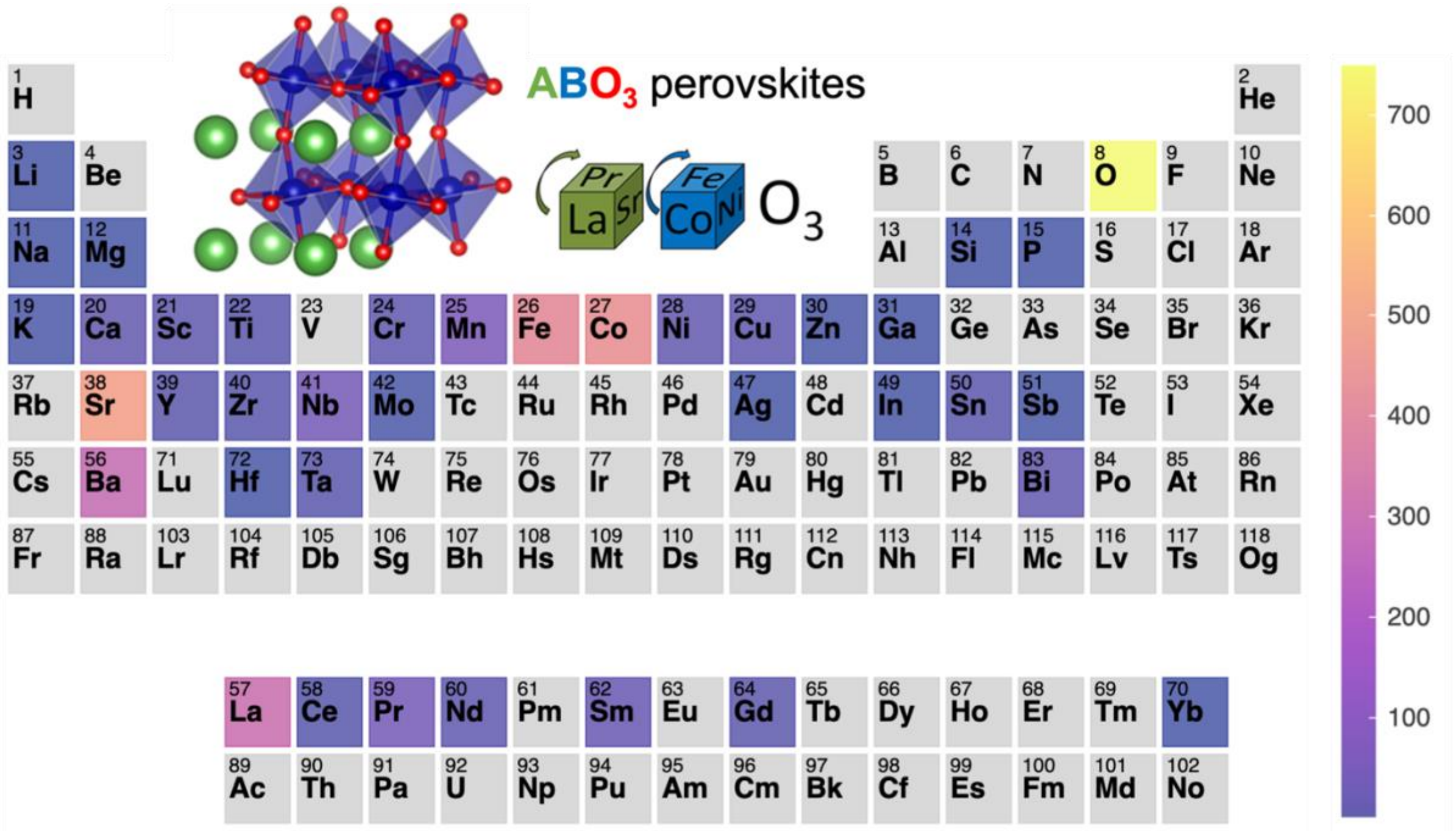

**Figure 1.** Periodic table heatmap summarizing the occurrences of different elements in the perovskite compositions comprising our catalytic property database.

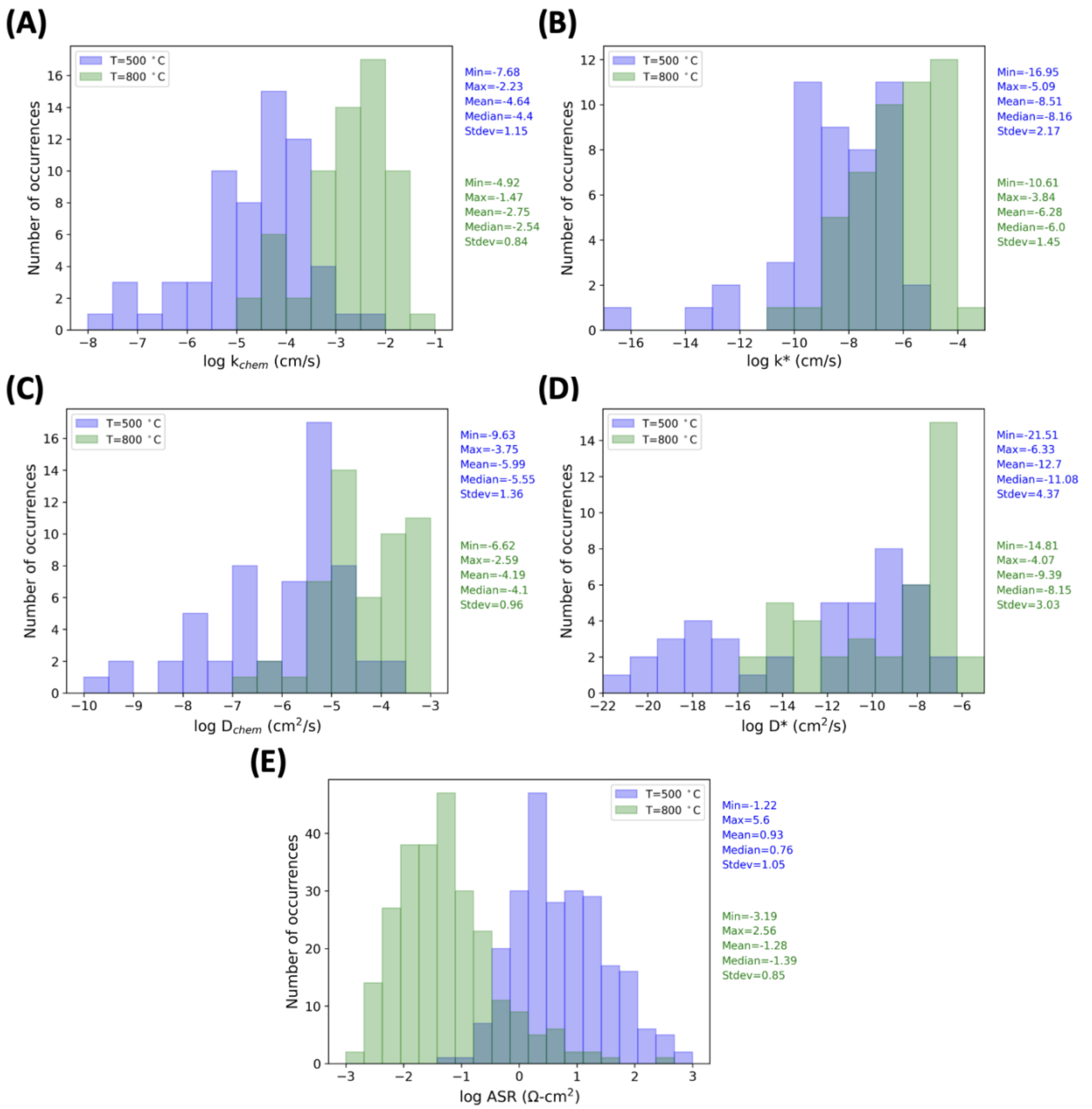

**Figure 2.** Histogram distributions of each perovskite catalytic property of interest from our collected database. In all plots, blue bars represent the distribution at 500 °C and green bars represent the distribution at 800 °C. (A): $k_{chem}$ distribution, (B): k* distribution, (C): $D_{chem}$ distribution, (D): D* distribution, (E): ASR distribution. Note that all values are plotted as the base-10 logarithm of the property value.

## 2.2. Assessment of O p-band center trends

In this section, we assess trends of our perovskite catalytic properties as a function of the DFT-calculated O p-band center (see **Methods** in **Section 4.1**). These assessments serve as a more detailed test of the O p-band center descriptor now that the perovskite catalytic property database has been significantly expanded compared to previous studies. Previous work by Lee et al.[46] formed correlations of k* versus O p-band center for 9 materials and ASR versus O p-band center for 7 materials, resulting in linear $R^2$ fit values of 0.87 and 0.99, respectively (note that LMO and LSM distinctly fall off the linear trend line of ASR versus O p-band in this example, and were not included in the fit). Subsequent work by Jacobs et al.[55] formed correlation of k* versus O p-band center for 21 materials with a linear $R^2$ fit of 0.86 and follow-on study of a subset of 9 k* measurements also had a linear $R^2$ fit of 0.87.[53] These previous studies examined at least a 3-5× smaller set of material than considered in this work, and only examined the trends at a single temperature.

**Figure 3** contains plots of perovskite catalytic properties from ECR ($k_{chem}$, $D_{chem}$), IEDP (k*, D*) and ASR values from EIS measurements as a function of the O p-band center calculated from DFT for each material. In **Figure 3**, the catalytic property trends are plotted at both 500 °C and 800 °C. Equivalent plots at 600 °C and 700 °C are provided in **Figure S2** of the **SI**. We observe from **Figure 3** that the quality of linear fit to the O p-band center is both property- and temperature-dependent. In general, the fit trends improve when considering data at higher temperatures. Based on the fit mean absolute error (MAE), when going from 500 °C to 800 °C the MAEs of fits to $k_{chem}$ (0.68 vs. 0.49 cm/s), k* (1.3 vs. 0.76 cm/s), $D_{chem}$ (0.72 vs. 0.56 $cm^2/s$), D* (2.68 vs. 1.79 $cm^2/s$) and ASR (0.69 vs. 0.57 Ohm-$cm^2$) all decrease. We speculate the improved fits at higher temperature are a result of less experimental uncertainty in the measured values at higher temperature, producing less spread in the data and producing improved fits. The MAE values of k* and D* are higher than their $k_{chem}$ and $D_{chem}$ analogues, however this is likely a result of the larger dynamic range of the k* and D* data relative to $k_{chem}$ and $D_{chem}$, because the reduced root mean square error (RMSE/σ) (σ= dataset standard deviation) values for all properties are qualitatively similar, in the range of 0.6-0.8, indicating from this metric that the fits are statistically of similar quality.

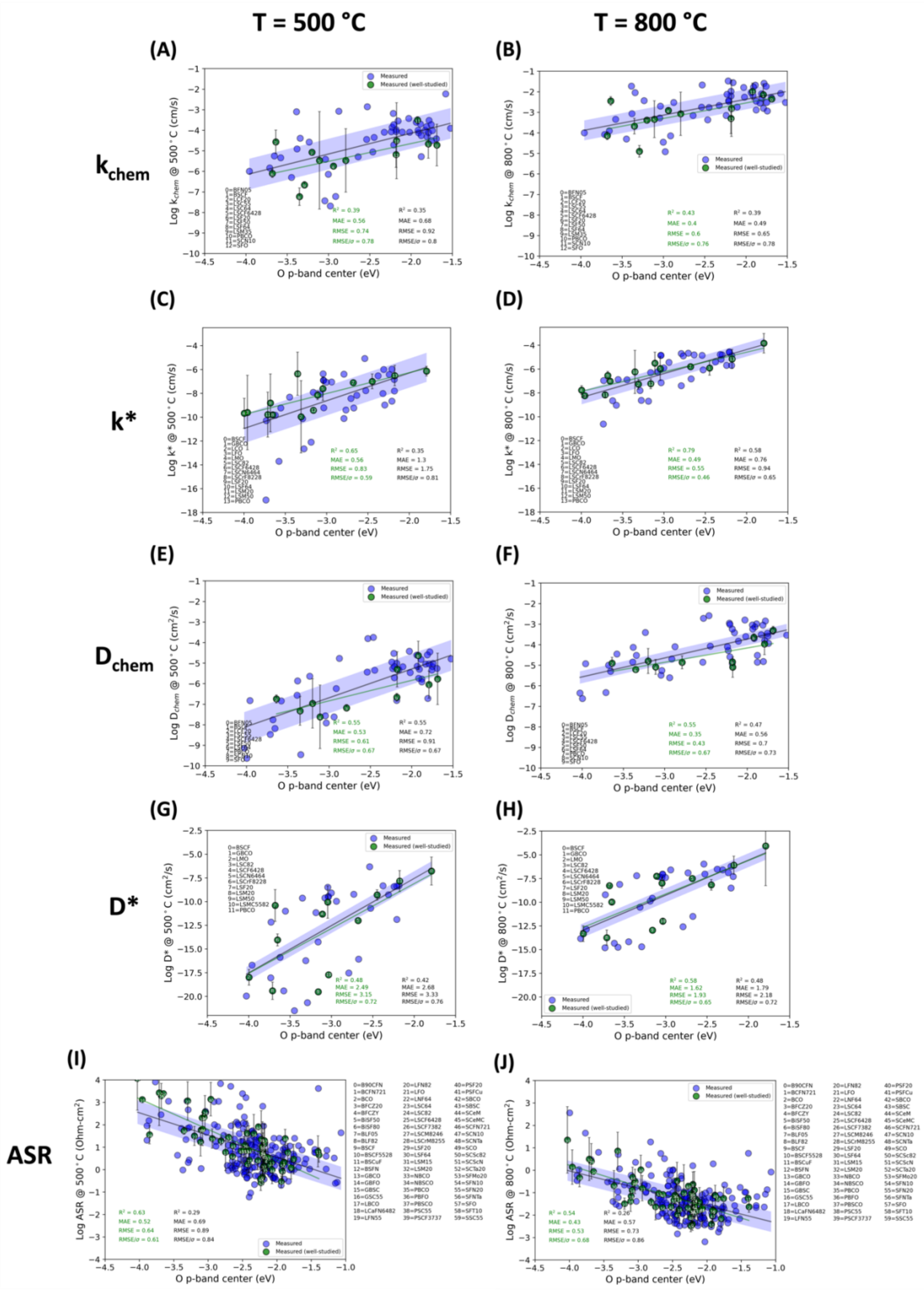

**Figure 3.** Trends of perovskite catalytic properties as a function of O p-band center. The left (right) column of plots is for T = 500 °C (T = 800 °C). The properties plotted are $k_{chem}$ (A, B), k* (C, D), $D_{chem}$ (E, F), D* (G, H), and ASR (I, J). The blue points and black fit line and statistics are for a linear fit to all the data. The green points and green fit line and statistics are for a linear fit to the subset of well-studied materials only (those materials with 2 or more measurements). The blue "x" points are outliers which were removed from the trend analysis. The blue shaded region is the average experimental error range, where the experimental error is calculated as the standard deviation of materials with multiple measurements. The average experimental error bars are 0.68 and 0.45 log units at 500 °C and 800 °C, respectively. For the ASR trend, ASR measurements of the same material with different electrolytes are averaged together. Data of the slopes and intercepts for all linear fits can be found in

The quality of the fits in **Figure 3** varies substantially when considering a fit to all materials (blue + green points) or just the subset of well-studied materials (green points), where here "well-studied" denotes materials with 2 or more independent measurements in the database. In **Figure 4**, the trend of improved fits both at higher temperatures and when considering only well-studied materials is shown more explicitly. In general, the well-studied materials have MAE smaller than fitting to all data points by about 10-50%, depending on the property and temperature. This result strongly suggests that a significant source of the errors in correlations with O p-band center is uncertainty in the experiments, where any given experimental data point may have an error bar of approximately 0.5-0.7 log units, depending on the temperature. We hypothesize that when a particular material is measured multiple times by different research groups, a more reliable average value emerges, and this causes the fits of O p-band center to have significantly less error. In **Figure 4** we alter the definition of "well-studied" to range from single measurements (the limiting case of considering all data in the database), up to the case where 5 or more independent measurements were conducted for a given material. Consistent with the above hypothesis, we find that model accuracy generally improves up to at least 3 or more measurements for a given material, as illustrated in **Figure 4F.** From **Figure 4F**, we can see that at 500 °C (800 °C), the MAE (in log units) of ASR is reduced from 0.58 to 0.38 (0.51 to 0.26) Ohm-cm$^2$, corresponding to percentage reductions of about 35% (50%), respectively.

The quality of the linear fits in this work are slightly lower, but similar to, reliable values reported in previous studies discussed above when one takes into account that the previous studies were generally focused on high temperatures (about 800 °C) and values that came from averaging multiple studies. Specifically, previous results found an $R^2$ for k* vs. p-band of 0.86-0.87[46,53,55] while the $R^2$ for the well-studied materials at 800 °C from this work are 0.79 (**Figure 3D**). This modest reduction is likely due to the very high quality and limited data used in the earlier studies, leading to good correlations and significant sampling uncertainty. Previous studies found an $R^2$ for ASR vs. p-band of 0.99[46], which is much higher that the 0.44 (**Figure 3J**) found in this work. However, the extremely high 0.99 value is almost certainly an overestimate, likely caused by that study having removed selected data under the assumption that it followed a different model, some model tuning (as this result was the first O p-band center study), and the very limited size of the remaining data (just 7 points).

Overall, the O p-band center descriptor is useful in that it can provide a qualitative estimate of numerous catalytic properties at a given temperature of interest, where performing a single, bulk DFT calculation is generally much easier and faster (approximately 10 hours of computer time and can be done in parallel, with just minutes of human time for a properly automated setup) than synthesizing a material and performing the needed property measurements (possibly weeks of human time, typically done in serial). However, one should not expect quantitatively accurate predictions of key perovskite catalytic properties, at least estimates with error less than approximately 0.5 log units. Furthermore, it is likely that DFT-based descriptor approaches like the O p-band center are likely to be partially or totally replaced by more accurate and faster data-centric machine learning approaches as databases grow.

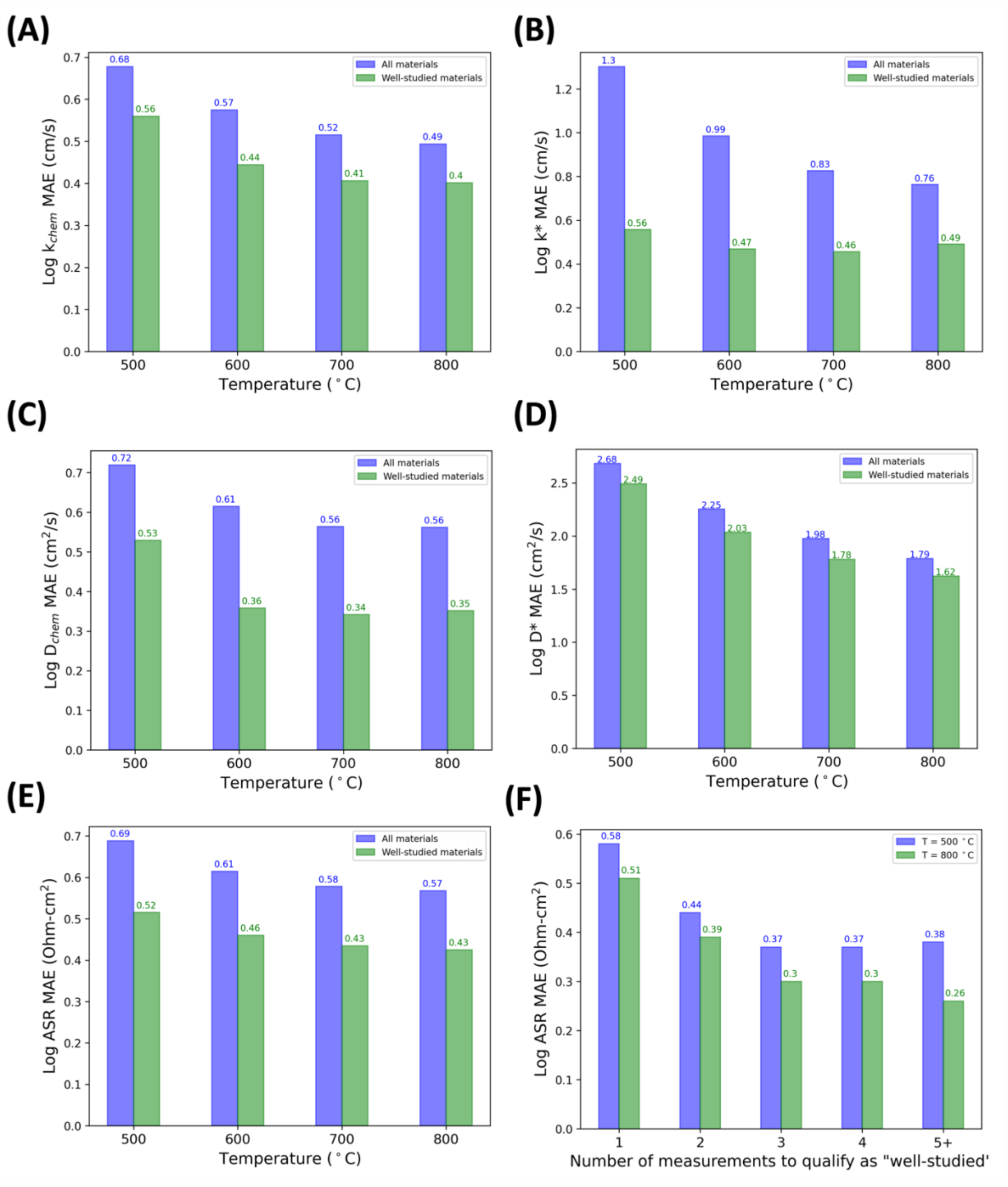

**Figure 4.** Bar charts summarizing the MAE values of linear fits to O p-band center of perovskite catalytic properties from ECR and IEDP experiments. In plots (A)-(E), the blue bars are MAE values of linear fits to all of the data, while the green bars are MAE values of linear fits to the subset of well-studied materials only (materials with 2 or more measurements). The properties are $k_{chem}$ (A), k* (B), $D_{chem}$ (C), D* (D), and ASR (E). Plot (F) further shows how the MAE on the ASR values varies when considering fits to materials containing increasing numbers of experimental measurements.

## 2.3. Summary of assessing relationships between catalytic properties

The database of perovskite catalytic properties developed in this work can be used to assess correlations between these properties proposed from previous studies. Such correlations can provide mechanistic insights and be used to fill in missing data with approximate values as a guide to material performance before performing a particular experiment or assessment of a result after experiment. In **Section S3** of the **SI**, we provide detailed discussion of assessing three inter-property relationships which have been proposed in the literature: First, the work of Kilner et al. showed there is a linear correlation between k* and D* for a set of $Sm_{1-x}Sr_xCoO_3$, $La_{1-x}Sr_xCoO_3$ and $La_{1-x}Sr_xMnO_3$ perovskites.[65] Our analysis at 800 °C shows very good linear correlation similar to the work of Kilner et al.,[65] but with a wider range of perovskite compositions here ($R^2$ = 0.75 for all materials, $R^2$=0.93 for well-studied materials). Second, Lee et al. and Jacobs et al. showed correlations of k* versus ASR for a limited set of materials.[46,55] This trend is particularly useful as it directly relates k*, which is a materials property of the perovskite electrode catalyst, to ASR, which is related to the perovskite materials property but also includes additional confounding factors of the overall electrode device, such as its microstructure, the electrolyte it is bonded with, and any interfacial structure or phases at the electrode/electrolyte interface that may form. For this relationship of k* versus ASR, our analysis at 800 °C results in an $R^2$=0.63, which increases to $R^2$=0.81 for well-studied materials. Third, seminal work from Adler, Lane and Steele formulated a detailed model relating ASR with k*, D*, and factors related to electrode microstructure, where this Adler-Lane-Steele[66] relationship stipulates that ASR is proportional to the inverse square root of the product of k* and D*. Similar with the other properties above, we find that this correlation persists but is improved when considering only well-studied materials, where at 800 °C we obtain $R^2$=0.68, increased slightly to $R^2$=0.7 for well-studied materials. Overall, all of these property correlations follow the same two general principles discussed at length in **Section 2.2**: namely, the correlations are more robust at higher temperature and when considering fits to well-studied materials only, and show linear correlations of modest quality likely useful for qualitative materials property prediction.

## 2.4. Using thermodynamic factors and property correlations to predict new data

The inter-property correlations discussed in **Section 2.3** and **Section S3** of the **SI** provide one way to predict new property values given the corresponding property value forming the correlation. Here, we describe another approach to predict new materials property data by using the thermodynamic factor, $\gamma$, which relates the tracer surface exchange and diffusion with their chemical analogues, i.e., $k_{chem} = \gamma k^*$ and $D_{chem} = \gamma D^*$.[27,67] In this section, we first compare the values of $\gamma$ obtained in three ways, then we use $\gamma$ and the correlation of k* versus ASR detailed in **Section S3** of the **SI** to obtain new predicted values of k* using measurements of $k_{chem}$ and ASR. Knowledge of $\gamma$ can thus be used to fill in missing data values (e.g., obtain k* if only $k_{chem}$ is known), thus saving experimental time and aiding in screening for high-performing materials. Here, values of $\gamma$ for a given material may be directly measured, which we refer to as "measured $\gamma$", the values of $\gamma$ may be extracted from separate measurements of either $k_{chem}/k^*$ or $D_{chem}/D^*$, which we refer to as "extracted $\gamma$", and, finally, $\gamma$ may be predicted by taking the ratio of the linear fit lines from $k_{chem}$ and k* versus O p-band center (or $D_{chem}$ and D* versus O p-band center), which we refer to as "predicted $\gamma$".

It is worth noting that extracting $\gamma$ from value pairs of $k_{chem}/k^*$ or $D_{chem}/D^*$ involves values taken from different research groups and different studies. Such a comparison may therefore add additional error as no two groups of people will make the exact same material or run a given experiment or analyze data in the same way. To our knowledge, there is only one case where $\gamma$ was measured for a material, along with its $k_{chem}$ and k* values in the same study.[33] This study from Bucher et al.[33] sought to explicitly test the relationships of $k_{chem} = \gamma k^*$ and $D_{chem} = \gamma D^*$ against measuring $\gamma$ using a precision thermobalance. They performed this test for BSCF. At T = 600 °C they measured $\gamma$ to be 125 +/- 2, and values of $k_{chem}$ and k* at 600 °C resulted in an extracted $\gamma$ of about 110, and values of $D_{chem}$ and D* at 600 °C resulted in an extracted $\gamma$ of 125, both of which showing essentially perfect agreement with the measured $\gamma$. In addition, a similar study from Saher et al. measured k* and $k_{chem}$ for LSCF and used $\gamma$ values published in a separate study.[68] Both of these studies verified that the relationship between the tracer and chemical properties and $\gamma$ is robust, and suggest that the differences of measured versus extracted $\gamma$ we show below in **Figure 5** are likely the result of differences of approach between various researchers.

**Figure 5** compares values of $\gamma$ obtained from extraction of $k_{chem}/k^*$ values, $D_{chem}/D^*$ values, and predicted using the ratios of their respective O p-band center trendlines from **Figure 3**. For the data in **Figure 5**, a temperature of T = 800 °C was used, because it resulted in the most experimentally measured $\gamma$ values being available for comparison. Separately, in **Figure S5** of the **SI**, we plot the $\gamma$ values extracted from $k_{chem}/k^*$ as a function of temperature, with experimentally measured points overlaid. For the analysis shown in **Figure 5**, we were restricted to the set of materials in our database for which all of $k_{chem}$, $k^*$, $D_{chem}$ and $D^*$ had been measured, yielding only 12 materials. We were also able to compare these extracted and predicted $\gamma$ values with those measured directly from experiment, of which data was available for 8 of the 12 materials. From the comparisons in **Figure 5**, we observe that the $\gamma$ values extracted from $k_{chem}/k^*$ measurements (dark green bars) are in closer agreement with the experimentally measured $\gamma$ values (red bars) than those extracted from $D_{chem}/D^*$ measurements (light green bars). More specifically, the MAE of $\gamma$ in log units between these two values is 0.46 versus 1.29 for extracting $\gamma$ from $k_{chem}/k^*$ and $D_{chem}/D^*$, respectively. In addition, the predicted $\gamma$ values from the O p-band center trends with $k_{chem}$ and $k^*$ also produce good agreement both with the extracted $\gamma$ values from $k_{chem}/k^*$ (dark purple bars) and the measured values (red bars) (MAE of 0.54 log units), while the predicted $\gamma$ from the trends of $D_{chem}$ and $D^*$ (light purple bars) result in higher errors with many overestimations of the true $\gamma$ values (MAE of 1.88 log units). The reason for the worse ability of using $D_{chem}$ and $D^*$ values to obtain $\gamma$ is not totally clear but may be in part due to the larger spread of $D^*$ values in experiment relative to $k^*$ values, and the relatively worse correlation of $D^*$ versus the O p-band center compared to the analogous $k^*$ trend.

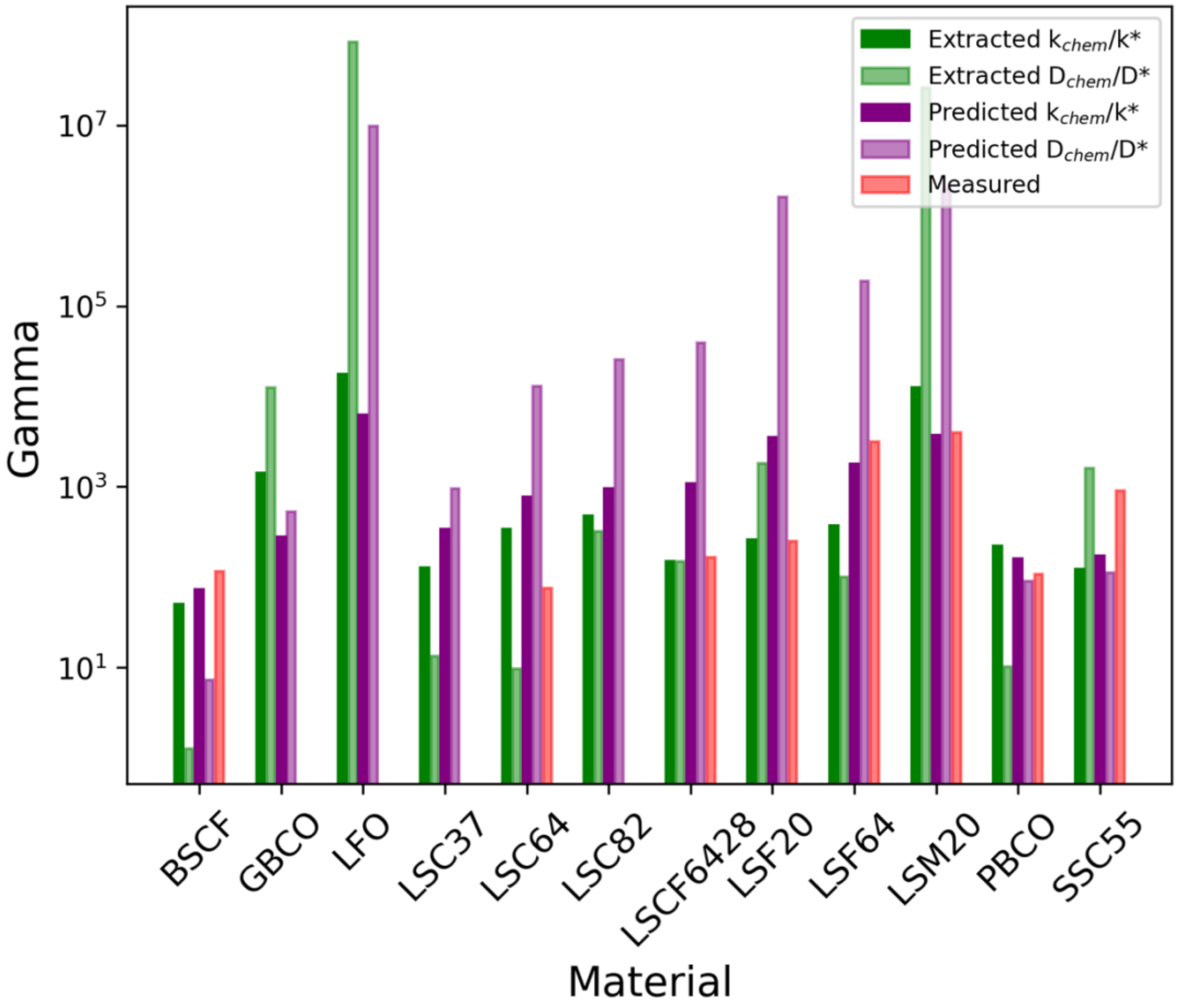

**Figure 5.** Bar chart summarizing thermodynamic factors ($\gamma$) extracted from perovskite catalytic properties database, predicted from linear fits of catalytic properties with O p-band center, and from direct experimental measurements. All of these data are for a temperature of T = 800 °C.

While the use of the predicted $\gamma$ from $k_{chem}$/k* has slightly higher error than if measured values of $k_{chem}$ and k* were used, this method offers an avenue to convert between $k_{chem}$ and k* values using only a single experimental measurement and the DFT-calculated O p-band center, as opposed to needing to perform two separate experiments. In **Figure 6**, we take the 45 $k_{chem}$ values in our database for which there is no corresponding k* value, and predict the corresponding k* values by using the predicted $\gamma$ value from the ratio of the O p-band center trend lines of $k_{chem}$ and k*. These 45 predicted k* values are shown as purple dots in **Figure 6**. In addition, we use the inter-property correlation of k* versus ASR from **Section S3** of the **SI** and predict k* values for the 235 materials for which ASR data is known but no k* value exists in the database. These 235 k* predictions are shown as red dots in **Figure 6** (they are also available as part of the data files provided as part of the **SI**). This scheme provides quick qualitative estimates of catalytic property data and can be useful for fleshing out existing data, making decisions on key experiments to perform, or screening of top-performing materials based on specific performance metrics. From our list of newly predicted k* values, the materials BiSFSb10, PSCF4628, BFZ90, and BiPBF (BiSFSb10, PSCF4628 are materials with low ASR in our database, while BFZ90 and BiPBF have high $k_{chem}$) have predicted log k* values of -5.12, -5.28, -2.88, and -4.03 cm/s at T = 500 °C, respectively, all of which are comparable or higher than the top k* material, LSC37 (see discussion in **Section S1** of the **SI**), with a log k* value of -5.09 cm/s at 500 °C.

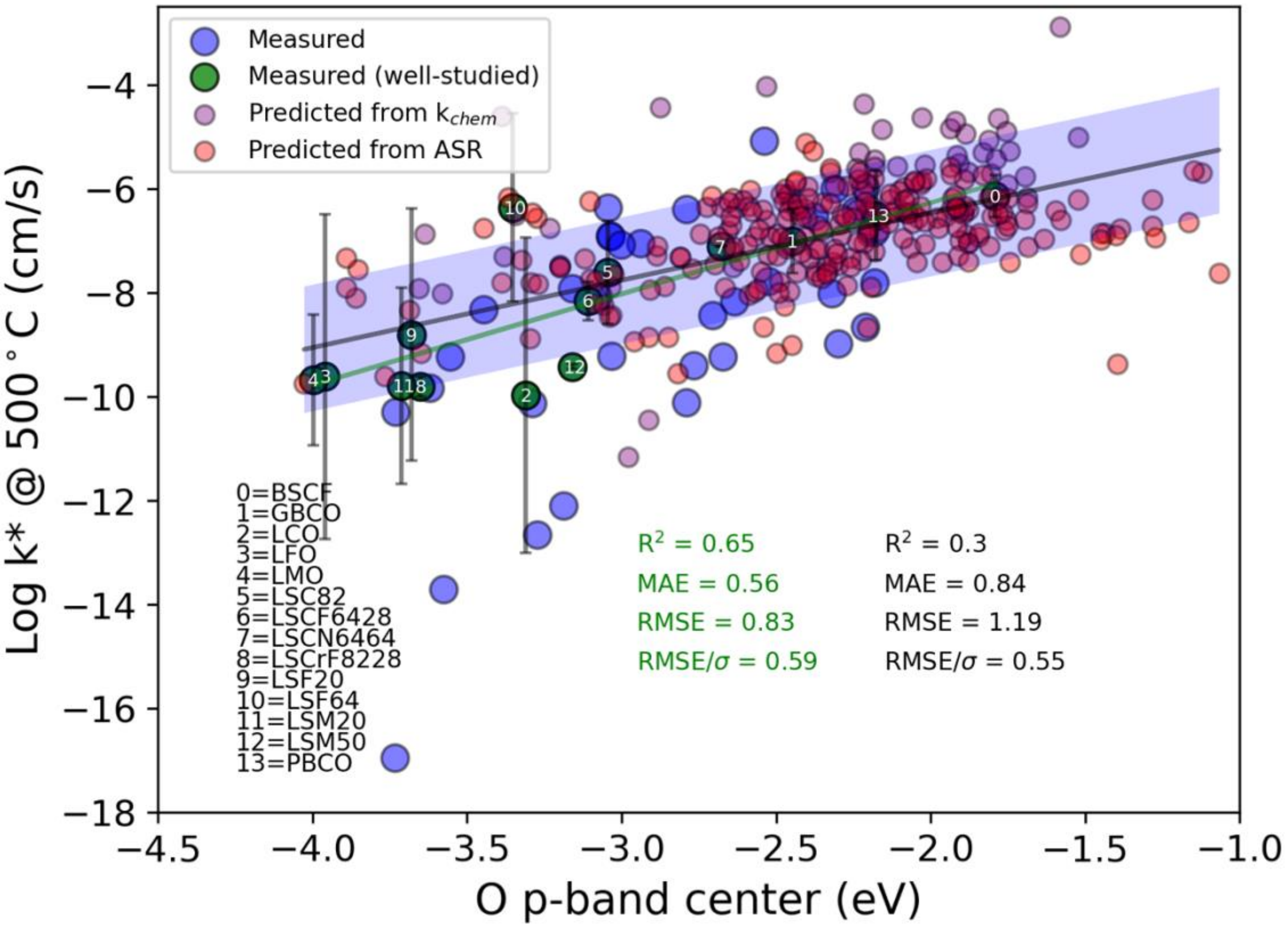

**Figure 6.** Trend of k* with O p-band center at 500 °C with newly predicted data. The blue and green data are measured k* values and well-studied measured k* values, respectively. The purple data are predicted k* values from measured $k_{chem}$ and predicted $\gamma$. The red data are predicted k* values from measured ASR values.

# 3. Summary and Conclusion

In this work, we constructed the largest database to-date of perovskite catalytic properties comprising oxygen surface exchange rates ($k_{chem}$ and k*), oxygen diffusivity ($D_{chem}$ and D*), and ASR data from ECR, IEDP, and EIS experiments. The database consists of nearly 750 measured property values spanning 299 unique perovskite materials from over 300 studies. By itself, the database is useful for examining typical value ranges at a given temperature, understanding typical ranges of experimental error from multiple independent measurements, flagging outlier data points as potentially problematic measurements, and listing top-performing materials based on each property. Using our database, we assess the ability of the O p-band center electronic structure descriptor, calculated using DFT methods, to correlate with $k_{chem}$, k*, $D_{chem}$, D* and ASR values. We generally find that there are modest linear trends of each property with the O p-band center when considering fits to the full database of materials. We find that the fits are more robust at higher temperature, and significantly improve when considering fits to the subset of materials which have multiple independent measurements. These findings suggest that the spread of property data is largely the result of experimental uncertainty, and subsequent measurements of additional materials will continue to improve the O p-band center correlations. We also use our database to assess inter-property correlations, such as k* versus D*, k* versus ASR, and the Adler-Lane-Steele formula relating ASR to k* and D*. We broadly find that using the O p-band center results in property prediction mean absolute errors on the order of 0.5 log units, depending on the specific property and temperature considered. Further, we use our inter-property correlations with information of thermodynamic factor values extracted from the data and predicted from our O p-band correlations to make predictions of k* for materials based on measured values of $k_{chem}$ and ASR, and list several potential fast oxygen surface exchange materials which have not previously been explored using IEDP methods.

# 4. Data and Methods

## 4.1. Density Functional Theory (DFT) calculations

All Density functional theory (DFT) calculations were conducted using the Vienna Ab Initio Simulation Package (VASP).[69] Materials were modeled using a planewave basis set and the Generalized Gradient Approximation (GGA) exchange and correlation functional with Perdew-Burke-Ernzerhof (PBE) pseudopotentials.[70] The Hubbard U correction was not used in the present work following the study from Jacobs et al.,[53] who found the O p-band center trends were more robust over a range of properties for GGA-PBE compared to PBE+U. Spin polarization was enabled for all calculations. A planewave cutoff energy of 500 eV was used for all calculations. Perovskite materials were simulated using a 2×2×2 supercell (40 atoms) and a 4×4×4 Monkhorst-Pack[71] k-point mesh. To more accurately capture the experimental stoichiometry, for select materials, a larger 2×2×4 supercell (80 atoms) was used with a 4×4×2 Monkhorst-Pack k-point mesh was used. All perovskite compositions were first initially fully relaxed (volume + ions) to obtain the correct equilibrium volume. Then, a second run was performed where the cell parameters were set to be pseudocubic, that is, lattice parameters $a = b = c$ but with the same equilibrium volume as obtained from the full relaxation run. For this second run, only ions were relaxed. The electronic density of states output from this second pseudocubic relaxation was used to calculate the O p-band center following previous studies.[46,53,55] All materials were simulated as fully stoichiometric with no oxygen vacancies. We acknowledge that these choices of simulating all materials as stoichiometric and pseudocubic is not physically rigorous for all systems, however doing so provides us a consistent set of calculations to more straightforwardly assess correlations between properties without introducing additional confounding factors from the simulation approach.

## 4.2. Database development

To extract perovskite catalytic data from the literature, each paper was manually evaluated and searched for data in the text, in tables, and in figures. Most of the data was found to reside in figures, where typically data of any of the properties of interest is plotted as a function of temperature. We manually extracted measured property values at two different temperatures, so that an activation energy barrier could be obtained, and the property value for a given material could then be scaled to any desired temperature. For the ASR data, we did not explicitly differentiate data which originated from symmetrical cell tests versus full button cell tests, but most of the data came from symmetrical cell tests. In addition to extracting the ASR values, the microstructure type of the electrode studied (porous versus dense) and the electrolyte material used were also documented. There were some instances of studies where a property value was reported at only a single temperature. For these cases, we impute a value for the activation barrier. If the material for which the activation barrier requires imputation is present in our database, the average value of the activation barrier of the other measurements was used. If the material is not in the database, but a closely related material is (e.g., LSM35 not present, but LSM20 and LSM50 were), then the average of the nearby compounds was used. We note while this is a subjective designation, it was only needed for these cases: LSM20 (LSM35 closest), LSM35 (LSM20 and LSM50 closest), LSF30 (LSF20 and LSF50 closest), PBFO (PBCO closest), LSF64 (LSF20 and LSF10 closest), and LSMSc (LSMSc20 closest). Finally, if the material and no compositionally close material is present, the average activation barrier of the entire database was used. A comparison of 165 activation barriers obtained from the literature data versus the corresponding activation barrier values calculated from data extracted from figures resulted in a mean absolute error of 0.07 eV in the activation barrier, which is roughly 5% of a typical activation barrier value. We note that this error in the activation barrier is about the same level of error one may obtain from a linear fit to the data experimentally. For example, activation barrier error values quoted from Bucher et al.[33] and Saher et al.[68] were in the range of 7-12 kJ, which is 0.07-0.12 eV. Therefore, we expect our manual data extraction scheme to yield reasonably accurate property values as a function of temperature, as the activation barriers have about the same level of error as observed in typical experimental fits. It is worth noting that an activation barrier error of 0.07 eV, which, for a typical barrier of 1.2 eV, would yield a property change of +/- 0.3 log units at 800 °C. As discussed throughout this work, we believe a main source of error in our correlation fits is the result of such experimental uncertainty. Errors within about 0.3 log units may be a useful rule of thumb for establishing a lower bound on the precision one may obtain from such correlations, as any further error reduction would be within the uncertainty of what is typically measured experimentally.

In an effort to properly credit all original authors for their work performing the materials property measurements for all data used in this work, below we provide in **Table 2** a detailed record of references for each perovskite composition and the property which was measured.

**Table 2.** Detailed summary of original references examined in this work to obtain the present database of perovskite catalytic properties.

| Material | Abbreviated composition | $k_{chem}$ | $D_{chem}$ | k* | D* | ASR |
|---|---|---|---|---|---|---|
| $Ba_{0.9}Co_{0.7}Fe_{0.2}Nb_{0.1}O_3$ | B90CFN | | | | | 72,73 |
| $Ba_2Bi_{0.1}Sc_{0.2}Co_{1.7}O_5$ | BBiScC | | | | | 74 |
| $Ba_{0.95}Ca_{0.05}FeO_3$ | BCaF | | | | | 75 |
| $Ba_{0.95}Ca_{0.05}Fe_{0.95}In_{0.05}O_3$ | BCaFIn | | | | | 75 |
| $BaCe_{0.5}Fe_{0.5}O_3$ | BCeF55 | 76 | | | | 76 |
| $Ba_{0.9}Co_{0.7}Fe_{0.2}Mo_{0.1}O_3$ | BCFMo | | | | | 77 |
| $BaCo_{0.6}Fe_{0.3}Nb_{0.1}O_3$ | BCFN631 | 78 | 78 | | | 78 |
| $BaCo_{0.7}Fe_{0.2}Nb_{0.1}O_3$ | BCFN721 | | | | | 72,73,79–81 |
| $BaCo_{0.7}Fe_{0.22}Sc_{0.08}O_3$ | BCFSc | 82 | 82 | | | 82 |
| $BaCo_{0.6}Fe_{0.2}Sn_{0.2}O_3$ | BCFSn20 | 11 | 11 | | | 11 |
| $BaCo_{0.6}Fe_{0.3}Sn_{0.1}O_3$ | BCFSn631 | 78 | 78 | | | 78 |
| $BaCo_{0.7}Fe_{0.2}Sn_{0.1}O_3$ | BCFSn721 | 83 | 83 | | | |
| $BaCo_{0.6}Fe_{0.2}Sn_{0.1}Y_{0.1}O_3$ | BCFSnY | 11 | 11 | | | 11 |
| $BaCo_{0.7}Fe_{0.22}Y_{0.08}O_3$ | BCFY | 84 | 84 | | | 84 |
| $BaCo_{0.6}Fe_{0.2}Y_{0.2}O_3$ | BCFY20 | 11 | 11 | | | 11 |
| $BaCo_{0.7}Fe_{0.2}Yb_{0.1}O_3$ | BCFYb | | | | | 85 |
| $BaCoO_3$ | BCO | | | | | 86,87 |
| $BaCo_{0.875}Sc_{0.125}O_3$ | BCSc12 | | | | | 87 |
| $BaCo_{0.75}Sc_{0.25}O_3$ | BCSc25 | 87 | 87 | | | 87 |
| $BaCo_{0.625}Sc_{0.375}O_3$ | BCSc37 | | | | | 87 |
| $BaFe_{0.9}Bi_{0.1}O_3$ | BFBi | | | | | 88 |
| $BaFe_{0.6}Co_{0.3}Ce_{0.1}O_3$ | BFCCe631 | | | | | 89 |
| $BaFe_{0.95}Ce_{0.05}O_3$ | BFCe05 | | | | | 90 |
| $BaFe_{0.85}Cu_{0.15}O_3$ | BFCu15 | | | | | 91 |
| $BaFe_{0.75}Cu_{0.25}O_3$ | BFCu25 | | | | | 91 |
| $BaFe_{0.375}Co_{0.375}Zr_{0.25}O_3$ | BFCZ20 | 63 | 63 | | | 92,93 |
| $BaCo_{0.2}Fe_{0.5}Zr_{0.3}O_3$ | BFCZ253 | | | | | 94 |
| $BaFe_{0.25}Co_{0.25}Zr_{0.5}O_3$ | BFCZ50 | 63 | 63 | | | |
| $BaCo_{0.5}Fe_{0.2}Zr_{0.3}O_3$ | BFCZ523 | | | | | 94 |
| $BaFe_{0.125}Co_{0.125}Zr_{0.75}O_3$ | BFCZ75 | 63 | 63 | | | |
| $BaFe_{0.4}Co_{0.4}Zr_{0.1}Y_{0.1}O_3$ | BFCZY | 95 | 95 | | | 37,95 |
| $BaFe_{0.95}In_{0.05}O_3$ | BFIn05 | | | | | 75 |
| $BaFe_{0.95}Nb_{0.05}O_3$ | BFN05 | 96,97 | 96,97 | | | 96 |
| $BaFe_{0.9}Nb_{0.1}O_3$ | BFN10 | 97 | 97 | | | |
| $BaFe_{0.8}Nb_{0.2}O_3$ | BFN20 | 97 | 97 | | | |
| $BaFe_{0.75}Ni_{0.25}O_3$ | BFNi82 | | | | | 98 |
| $BaFeO_3$ | BFO | | | | | 86 |
| $BaFe_{0.95}Sn_{0.05}O_3$ | BFSn05 | | | | | 99 |
| $BaFe_{0.8}Sn_{0.2}O_3$ | BFSn82 | | | | | 100 |
| $BaFe_{0.3}Sn_{0.2}Bi_{0.5}O_3$ | BFSnBi325 | | | | | 100 |
| $BaFe_{0.5}Sn_{0.2}Bi_{0.3}O_3$ | BFSnBi523 | | | | | 100 |
| $BaFe_{0.7}Sn_{0.2}Bi_{0.1}O_3$ | BFSnBi721 | | | | | 100 |
| $BaFe_{0.95}Zr_{0.05}O_3$ | BFZ05 | | | | | 90 |
| $BaFe_{0.1}Zr_{0.9}O_3$ | BFZ90 | 101 | | | | |
| $BaFe_{0.8}Zn_{0.1}Bi_{0.1}O_3$ | BFZnBi | | | | | 88 |
| $Ba_{0.95}Gd_{0.05}FeO_3$ | BGF05 | | | | | 90 |
| $Bi_{0.7}Pr_{0.1}Ba_{0.2}FeO_3$ | BiPBF | 102 | 102 | | | 102 |
| $Bi_{0.5}Sr_{0.5}Fe_{0.8}Co_{0.2}O_3$ | BiSCF | | | | | 103 |
| $Bi_{0.5}Sr_{0.5}FeO_3$ | BiSF50 | 104,105 | 104,105 | | | 104–108 |
| $Bi_{0.2}Sr_{0.8}FeO_3$ | BiSF80 | | | | | 106,107 |
| $Bi_{0.5}Sr_{0.5}Fe_{0.8}Cu_{0.2}O_3$ | BiSFCu | | | | | 109 |
| $Bi_{0.5}Sr_{0.5}Fe_{0.9}Nb_{0.1}O_3$ | BiSFN10 | | | | | 110 |
| $Bi_{0.5}Sr_{0.5}Fe_{0.9}Sb_{0.1}O_3$ | BiSFSb10 | | | | | 111 |
| $Bi_{0.5}Sr_{0.5}Fe_{0.9}Ti_{0.1}O_3$ | BiSFT10 | | | | | 112 |
| $Bi_{0.5}Sr_{0.5}Fe_{0.8}Ti_{0.2}O_3$ | BiSFT20 | | | | | 112 |
| $Bi_{0.5}Sr_{0.5}Fe_{0.9}Ta_{0.1}O_3$ | BiSFTa10 | | | | | 108 |
| $Ba_{0.95}La_{0.05}FeO_3$ | BLF05 | 113 | 113 | | | 90,113–115 |
| $Ba_{0.6}La_{0.4}FeO_3$ | BLF64 | | | | | 116 |
| $Ba_{0.8}La_{0.2}FeO_3$ | BLF82 | | | | | 116,117 |
| $Ba_{0.9}La_{0.1}FeO_3$ | BLF91 | | | | | 116 |
| $Ba_{0.95}La_{0.05}Fe_{0.95}P_{0.05}O_3$ | BLFP | | | | | 114 |
| $(Ba_{0.5}Sr_{0.5})_{0.8}Co_{0.8}Fe_{0.2}O_3$ | BS0.8CF | | | | | 118 |
| $Ba_{0.5}Sr_{0.5}Co_{0.8}Fe_{0.2}O_3$ | BSCF | 32,33,63,87,119–123 | 32,33,63,87,119,120,122,123 | 30,31,33 | 30,31 | 29,37,84,86,87,118,124–133 |
| $Ba_{0.2}Sr_{0.8}Co_{0.8}Fe_{0.2}O_3$ | BSCF2882 | | | | | 134 |
| $Ba_{0.5}Sr_{0.5}Co_{0.2}Fe_{0.8}O_3$ | BSCF5528 | | | | | 130,135 |
| $Ba_{0.5}Sr_{0.5}Co_{0.4}Fe_{0.6}O_3$ | BSCF5546 | | | | | 130,136 |
| $Ba_{0.5}Sr_{0.5}Co_{0.7}Fe_{0.2}Ni_{0.1}O_3$ | BSCFN | 136 | 136 | | | 136 |
| $Ba_{0.9}Sr_{0.1}Co_{0.9}In_{0.1}O_3$ | BSCIn | | | | | 137 |
| $Ba_{0.6}Sr_{0.4}Co_{0.9}Nb_{0.1}O_3$ | BSCN641 | 138 | 138 | | | 138 |
| $Ba_{0.5}Sr_{0.5}Cu_{0.2}Fe_{0.8}O_3$ | BSCuF | | | | | 130,139 |
| $Ba_{0.5}Sr_{0.5}Cu_{0.4}Fe_{0.6}O_3$ | BSCuF5546 | | | | | 130 |
| $Ba_{0.5}Sr_{0.5}Co_{0.1}Zn_{0.1}Fe_{0.8}O_3$ | BSCZnF | | | | | 135 |
| $Ba_{0.5}Sr_{0.5}FeO_3$ | BSF | | | | | 140 |
| $Ba_{0.5}Sr_{0.5}Fe_{0.8}Cu_{0.1}Ti_{0.1}O_3$ | BSFCuT | 141 | 141 | | | 141 |
| $Ba_{0.5}Sr_{0.5}Fe_{0.9}Mo_{0.1}O_3$ | BSFMo10 | | | | | 142 |
| $Ba_{0.5}Sr_{0.5}Fe_{0.9}Nb_{0.1}O_3$ | BSFN | | | | | 143,144 |
| $Ba_{0.5}Sr_{0.5}Fe_{0.9}Sb_{0.1}O_3$ | BSFSb10 | | | | | 140 |
| $Ba_{0.5}Sr_{0.5}Fe_{0.8}Zn_{0.2}O_3$ | BSFZn | | | | | 145 |
| $Ba_{0.4}Sr_{0.4}La_{0.2}FeO_3$ | BSLF | | | | | 146 |
| $Ba_{0.4}Sr_{0.4}La_{0.2}Fe_{0.9}Mn_{0.1}O_3$ | BSLFM10 | | | | | 146 |
| $Ba_{0.4}Sr_{0.4}La_{0.2}Fe_{0.9}Mn_{0.2}O_3$ | BSLFM20 | | | | | 146 |
| $Ba_{0.95}Sm_{0.05}FeO_3$ | BSmF05 | | | | | 90 |
| $Ba_{0.5}Sr_{0.5}Zn_{0.2}Fe_{0.8}O_3$ | BSZnF | | | | | 135 |
| $CaMn_{0.9}Nb_{0.1}O_3$ | CaMN91 | | | | | 147 |
| $CaMnO_3$ | CaMO | | | | | 147 |
| $GdBaCo_2O_5$ | GBCO | 148 | 148 | 34,149,150 | 34,149,150 | 34,127,151–153 |
| $GdBaFe_2O_5$ | GBFO | | | | | 154,155 |
| $GdBaMnFeO_5$ | GBMF | | | | | 156 |
| $Gd_{0.5}Ba_{0.5}Mn_{0.5}Fe_{0.5}O_3$ | GBMF5555 | | | | | 156 |
| $GdBa_{0.5}Sr_{0.5}Co_2O_5$ | GBSC | | | | | 157,158 |
| $GdBa_{0.5}Sr_{0.5}CoFeO_5$ | GBSCF | | | | | 157 |
| $Gd_{0.5}Sr_{0.5}CoO_3$ | GSC55 | | | | | 159,160 |
| $Gd_{0.8}Sr_{0.2}CoO_3$ | GSC82 | | | | | 161 |
| $La_{0.5}Ba_{0.5}CoO_3$ | LBC50 | | | | | 162 |
| $La_{0.4}Ba_{0.6}CoO_3$ | LBC60 | | | | | 163 |
| $La_{0.1}Ba_{0.9}Co_{0.4}Fe_{0.6}O_3$ | LBCF1946 | | | | | 164 |
| $La_{0.3}Ba_{0.7}Co_{0.6}Fe_{0.4}O_3$ | LBCF3764 | | | | | 164 |
| $La_{0.6}Ba_{0.4}Co_{0.2}Fe_{0.8}O_3$ | LBCF6428 | | | | | 129 |
| $LaBaCo_2O_5$ | LBCO | | | | | 127,162,165 |
| $LaBaFe_2O_5$ | LBFO | | | | | 154 |
| $La_{0.4}Ba_{0.6}Fe_{0.8}Zn_{0.2}O_3$ | LBFZn4682 | | | | | 166 |
| $La_{0.4}Bi_{0.4}Sr_{0.2}FeO_3$ | LBiSF | | | | | 167 |
| $La_{0.6}Ca_{0.4}Co_{0.2}Fe_{0.8}O_3$ | LCaCF6428 | | | 66 | 66 | |
| $La_{0.6}Ca_{0.4}Co_{0.8}Fe_{0.2}O_3$ | LCaCF6482 | | | 66 | 66 | |
| $La_{0.88}Ca_{0.12}CrO_3$ | LCaCr12 | | | 168 | 168 | |
| $La_{0.78}Ca_{0.22}CrO_3$ | LCaCr22 | | | 168 | 168 | 169 |
| $La_{0.7}Ca_{0.33}CrO_3$ | LCaCr35 | 170 | | 168 | 168 | |
| $La_{0.6}Ca_{0.4}Fe_{0.8}Ni_{0.2}O_3$ | LCaFN6482 | | | | | 171,172 |
| $La_{0.6}Ca_{0.4}Co_{0.8}Fe_{0.2}O_3$ | LCCF82 | | | 173 | 173 | |
| $La_{0.9}Ca_{0.1}FeO_3$ | LCF10 | 174 | 174 | | | |
| $La_{0.8}Ca_{0.2}FeO3$ | LCF20 | 175,176 | 175,176 | | | |
| $La_{0.9}Ca_{0.1}Fe_{0.9}Nb_{0.1}O_3$ | LCFN10 | | | | | 177 |
| $LaCo_{0.4}Ni_{0.4}Cu_{0.2}O_3$ | LCNCu | | | | | 178 |
| $LaCoO_3$ | LCO | | | 179,180 | 179 | 181 |
| $LaCrO_3$ | LCrO | | | | | 169 |
| $LaFe_{0.5}Ni_{0.5}O_3$ | LFN55 | | | | | 182 |
| $LaFe_{0.8}Ni_{0.2}O_3$ | LFN82 | | | | | 182 |
| $LaFeO_3$ | LFO | 86 | 86 | 180,183 | 183 | 86,182 |
| $LaMnO_3$ | LMO | | 184 | 25,180,185 | 25,185 | 184 |
| $LaNi_{0.6}Fe_{0.4}O_3$ | LNF64 | | | | | 186,187 |
| $(La_{0.75}Sr_{0.25})_{0.95}Cr_{0.5}Mn_{0.5}O_3$ | LS0.95CrM | | | 188 | 188 | |
| $La_{0.7}Bi_{0.1}Sr_{0.2}FeO_3$ | LSBiF12 | 189 | 189 | | | |
| $La_{0.6}Bi_{0.2}Sr_{0.2}FeO_3$ | LSBiF22 | 189 | 189 | | | |
| $La_{0.4}Bi_{0.4}Sr_{0.2}FeO_3$ | LSBiF42 | 189 | 189 | | | |
| $La_{0.3}Sr_{0.7}CoO_3$ | LSC37 | 190 | 190 | 191 | 191 | |
| $La_{0.5}Sr_{0.5}CoO_3$ | LSC55 | 190,192–195 | 190,192,193 | 196 | | |
| $La_{0.6}Sr_{0.4}CoO_3$ | LSC64 | 192,197 | 192 | 28 | 28 | 198–201 |
| $La_{0.8}Sr_{0.2}CoO_3$ | LSC82 | 190 | 190 | 26,66,173,180,202 | 26,66,173,202 | 161,200,203,204 |
| $La_{0.9}Sr_{0.1}CoO_3$ | LSC91 | | | 205 | 205 | |
| $La_{0.5}Sr_{0.5}Co_{0.8}Fe_{0.2}O_3$ | LSCF5582 | | | | | 126 |
| $La_{0.6}Sr_{0.4}Co_{0.2}Fe_{0.8}O_3$ | LSCF6428 | 27,67,68,119,123,206–210 | 27,67,68,119,123,206–210 | 27,68,180,206,211,212 | 27,68,180,206,211,212 | 63,130,131,159–161,212–219 |
| $La_{0.6}Sr_{0.4}Co_{0.8}Fe_{0.2}O_3$ | LSCF6482 | 220 | | 221 | 221 | |
| $La_{0.7}Sr_{0.3}Co_{0.8}Fe_{0.2}O_3$ | LSCF7382 | | | | | 203 |
| $La_{0.4}Sr_{0.6}Co_{0.2}Fe_{0.7}Nb_{0.1}O_3$ | LSCFN | | | | | 222 |
| $La_{0.8}Sr_{0.2}Co_{0.4}Mn_{0.6}O_3$ | LSCM8246 | | | | | 204 |
| $La_{0.1}Sr_{0.9}Co_{0.85}Nb_{0.15}O_3$ | LSCN | | | | | 223 |
| $La_{0.4}Sr_{0.6}Co_{0.6}Ni_{0.4}O_3$ | LSCN4664 | | | 224 | 224 | |
| $La_{0.4}Sr_{0.6}Co_{0.8}Ni_{0.2}O_3$ | LSCN4682 | | | 224 | 224 | |
| $La_{0.6}Sr_{0.4}Co_{0.4}Ni_{0.6}O_3$ | LSCN6446 | | | 66 | 66 | |
| $La_{0.6}Sr_{0.4}Co_{0.6}Ni_{0.4}O_3$ | LSCN6464 | | | 66 | 66 | |
| $La_{0.6}Sr_{0.4}Co_{0.8}Ni_{0.2}O_3$ | LSCN6482 | | | | 224 | |
| $La_{0.8}Sr_{0.2}Co_{0.8}Ni_{0.2}O_3$ | LSCN8282 | 225 | 225 | | | 225 |
| $La_{0.2}Sr_{0.8}Cr_{0.2}Fe_{0.8}O_3$ | LSCrF2828 | 226 | 226 | | | |
| $La_{0.3}Sr_{0.7}Cr_{0.3}Fe_{0.7}O_3$ | LSCrF3737 | | | | | 227 |
| $La_{0.4}Sr_{0.6}Cr_{0.2}Fe_{0.8}O_3$ | LSCrF4628 | | | 228 | 228 | |
| $La_{0.6}Sr_{0.4}Cr_{0.2}Fe_{0.8}O_3$ | LSCrF6428 | | | 228 | 228 | |
| $La_{0.8}Sr_{0.2}Cr_{0.2}Fe_{0.8}O_3$ | LSCrF8228 | | | 228,229 | 228,229 | |
| $La_{0.75}Sr_{0.25}Cr_{0.5}Mn_{0.5}O_3$ | LSCrM8255 | | | | | 230,231 |
| $La_{0.7}Sr_{0.3}CuO_3$ | LSCu73 | | | | | 232 |
| $La_{0.6}Sr_{0.4}Cu_{0.2}Fe_{0.8}O_3$ | LSCuF6428 | | | | | 233 |
| $La_{0.7}Sr_{0.3}Cu_{0.4}Fe_{0.6}O_3$ | LSCuF7346 | | | | | 232 |
| $La_{0.8}Sr_{0.2}Cu_{0.2}Fe_{0.8}O_3$ | LSCuF8228 | | | | | 219 |
| $La_{0.9}Sr_{0.1}FeO_3$ | LSF10 | | | 205 | 205 | |
| $La_{0.8}Sr_{0.2}FeO_3$ | LSF20 | 189,234 | 189 | 205 | 205 | 161,235 |
| $La_{0.7}Sr_{0.3}FeO_3$ | LSF30 | 210 | 210 | | | |
| $La_{0.5}Sr_{0.5}FeO_3$ | LSF50 | 105,226 | 105,226 | | | 105 |
| $La_{0.6}Sr_{0.4}FeO_3$ | LSF64 | 67,236 | 67,236 | 180,205 | 205 | 214,235 |
| $La_{0.5}Sr_{0.5}Fe_{0.8}Cu_{0.2}O_3$ | LSFCu | | | | | 237 |
| $La_{0.5}Sr_{0.5}Fe_{0.8}Ga_{0.2}O_3$ | LSFG82 | 238 | | | | |
| $La_{0.6}Sr_{0.4}Fe_{0.8}Mn_{0.2}O_3$ | LSFM6482 | 239 | 239 | | | 239 |
| $La_{0.5}Sr_{0.5}Fe_{0.9}Nb_{0.1}O_3$ | LSFN10 | | | | | 240 |
| $La_{0.85}Sr_{0.15}MnO_3$ | LSM15 | | 184 | | 241 | 184 |
| $La_{0.8}Sr_{0.2}MnO_3$ | LSM20 | 234 | 242 | 23,24,26,180,202 | 23,24,26,202,241 | 86,125,161,186,201,230,243–249 |
| $La_{0.7}Sr_{0.3}MnO_3$ | LSM35 | 250–252 | | 173 | 173 | 253 |
| $La_{0.5}Sr_{0.5}MnO_3$ | LSM50 | | | 66,173,196 | 66,173 | |
| $La_{0.5}Sr_{0.5}Mn_{0.8}Co_{0.2}O_3$ | LSMC5582 | | | 173 | 173,196 | |
| $La_{0.8}Sr_{0.2}Mn_{0.5}Co_{0.5}O_3$ | LSMC8255 | | | 196 | 196 | |
| $La_{0.8}Sr_{0.2}Mn_{0.8}Co_{0.2}O_3$ | LSMC8282 | | | 196 | | |
| $La_{0.8}Sr_{0.2}Mn_{0.8}Cu_{0.2}O_3$ | LSMCu | | | | | 254 |
| $La_{0.8}Sr_{0.2}Mn_{0.5}Fe_{0.5}O_3$ | LSMF8255 | | | 247 | 247 | |
| $La_{0.7}Sr_{0.3}Mn_{0.9}Mg_{0.1}O_3$ | LSMMg10 | | | | | 253 |
| $La_{0.7}Sr_{0.3}Mn_{0.8}Mg_{0.2}O_3$ | LSMMg20 | | | | | 253 |
| $La_{0.8}Sr_{0.2}Mn_{0.9}Sc_{0.1}O_3$ | LSMSc | | | | | 249 |
| $La_{0.8}Sr_{0.2}Sc_{0.2}Mn_{0.8}O_3$ | LSMSc20 | | | | | 255 |
| $La_{0.4}Sr_{0.6}Ni_{0.2}Fe_{0.8}O_3$ | LSNF4628 | | | | | 256 |
| $La_{0.3}Sr_{0.7}Ti_{0.3}Fe_{0.7}O_3$ | LSTF | | | | | 257 |
| $NdBa_{0.75}Ca_{0.25}Co_2O_5$ | NBCaCO | | | | | 258 |
| $NdBa_{0.5}Sr_{0.5}Co_{1.5}Fe_{0.5}O_5$ | NBCF1.5 | | | | | 259 |
| $NdBaCoMnO_5$ | NBCM | | | | | 260 |
| $NdBaCo_{1.5}Mn_{0.5}O_5$ | NBCM1.5 | | | | | 260 |
| $NdBaCo_2O_5$ | NBCO | | | | | 127,258,260–263 |
| $NdBaCu_2O_5$ | NBCuO | | | | | 264 |
| $NdBaFe_2O_5$ | NBFO | | | | | 154 |
| $NdBaMn_2O_5$ | NBMO | | | | | 260 |
| $NdBa_{0.5}Sr_{0.5}Co_{1.5}Mn_{0.5}O_5$ | NBSCMO | | | | | 265 |
| $NdBa_{0.5}Sr_{0.5}Co_2O_5$ | NBSCO | | | | | 263,265–267 |
| $NdBa_{0.5}Sr_{0.5}Cu_2O_5$ | NBSCuO | | | | | 264 |
| $NdCoO_3$ | NCO | | | | | 159 |
| $Nd_{0.5}Sr_{0.5}CoO_3$ | NSC55 | | | | | 159 |
| $Nd_{0.5}Sr_{0.5}Co_{0.5}Fe_{0.5}O_3$ | NSCF | | | | | 268 |
| $NdSrCo_2O_5$ | NSCO | | | | | 261 |
| $Nd_{0.5}Sr_{0.5}Fe_{0.8}Cu_{0.2}O_3$ | NSFCu | | | | | 237 |
| $PrBa_{0.8}Ca_{0.2}Co_{1.5}Fe_{0.5}O_5$ | PBCCFO | | | | | 269 |
| $PrBa_{0.8}Ca_{0.2}Co_2O_5$ | PBCCO | | | | | 269 |
| $PrBaCoFeO_5$ | PBCF | | | | | 270 |
| $PrBaCo_{1.5}Fe_{0.5}O_5$ | PBCF1.5 | | | | | 270 |
| $PrBaCo_2O_5$ | PBCO | 270–274 | 270,273 | 36,150,271,273,275 | 36,150,273,275 | 35,127,131,272,273,276–280 |
| $PrBaCo_{1.9}Sc_{0.1}O_5$ | PBCSc | | | | | 278 |
| $PrBaFe_2O_5$ | PBFO | 281 | 281 | | | 154,282 |
| $PrBa(Fe_{0.8}Sc_{0.2})_2O_5$ | PBFSc | | | | | 283 |
| $PrBaMn_{1.5}Fe_{0.5}O_5$ | PBMF1.5 | | | | | 284 |
| $PrBaMn_2O_5$ | PBMO | | | | | 284 |
| $PrBa_{0.5}Sr_{0.25}Ca_{0.25}Co_2O_5$ | PBSCaCO | | | | | 285 |
| $PrBa_{0.5}Sr_{0.5}CoCuO_5$ | PBSCCuO | | | | | 286 |
| $PrBa_{0.5}Sr_{0.5}CoFeO_5$ | PBSCFO | | | | | 287 |
| $PrBa_{0.5}Sr_{0.5}Co_2O_5$ | PBSCO | | | | | 158,280,285–288 |
| $Pr_{0.8}Ca_{0.2}BaCo_2O_5$ | PCBCO | | | | | 289 |
| $Pr_{0.8}Ca_{0.2}FeO_3$ | PCF20 | 175 | 175 | | | |
| $PrNi_{0.6}Co_{0.4}O_3$ | PNC64 | | | | | 290 |
| $Pr_{0.3}Sr_{0.7}CoO_3$ | PSC37 | | | | | 159 |
| $Pr_{0.5}Sr_{0.5}CoO_3$ | PSC55 | | | | | 159–161 |
| $Pr_{0.8}Sr_{0.2}CoO_3$ | PSC82 | | | | | 161 |
| $Pr_{0.3}Sr_{0.7}Co_{0.3}Fe_{0.7}O_3$ | PSCF3737 | | | | | 159,291 |
| $Pr_{0.4}Sr_{0.6}Co_{0.2}Fe_{0.8}O_3$ | PSCF4628 | | | | | 292 |
| $Pr_{0.6}Sr_{0.4}Co_{0.2}Fe_{0.8}O_3$ | PSCF6428 | | | | | 293 |
| $Pr_{0.4}Sr_{0.6}Co_{0.16}Fe_{0.64}Mo_{0.2}O_3$ | PSCFM20 | | | | | 292 |
| $Pr_{0.4}Sr_{0.6}Co_{0.2}Fe_{0.7}Nb_{0.1}O_3$ | PSCFN10 | | | | | 294 |
| $Pr_{0.6}Sr_{0.4}Co_{0.18}Fe_{0.72}Sc_{0.1}O_3$ | PSCFSc10 | | | | | 293 |
| $PrSrCo_2O_5$ | PSCO | | | | | 280 |
| $Pr_{0.8}Sr_{0.2}FeO_3$ | PSF20 | | | | | 161,295 |
| $Pr_{0.5}Sr_{0.5}Fe_{0.8}Cu_{0.2}O_3$ | PSFCu | | | | | 237,296 |
| $Pr_{0.7}Sr_{0.3}Fe_{0.8}Ni_{0.2}O_3$ | PSFN7382 | | | | | 297 |
| $Sr_{0.95}Co_{0.8}Nb_{0.1}Ta_{0.1}O_3$ | S95CNTa | | | | | 298 |
| $Sr_{0.95}Ag_{0.05}Nb_{0.1}Co_{0.9}O_3$ | SAgCN | 299 | | | | 299 |
| $SmBaCo_2O_5$ | SBCO | | | 150 | 150 | 127,300–302 |
| $SmBaFe_2O_5$ | SBFO | | | | | 154 |
| $SmBa_{0.5}Sr_{0.5}Co_2O_5$ | SBSC | | | | | 158,302–305 |
| $SmBa_{0.5}Sr_{0.5}Co_{1.5}Cu_{0.5}O_5$ | SBSCCu | | | | | 303 |
| $SmBa_{0.5}Sr_{0.5}Cu_2O_5$ | SBSCu | | | | | 306 |
| $SmBa_{0.5}Sr_{0.5}CuFeO_5$ | SBSCuF | | | | | 306 |
| $Sr_{0.9}Ce_{0.1}CoO_3$ | SCeC10 | | | | | 161 |
| $Sr_{0.9}Ce_{0.1}Co_{0.9}Nb_{0.1}O_3$ | SCeCN10 | | | | | 307 |
| $Sr_{0.9}Ce_{0.1}FeO_3$ | SCeF10 | | | | | 161 |
| $Sr_{0.9}Ce_{0.1}Fe_{0.8}Ni_{0.2}O_3$ | SCeFN82 | | | | | 308 |
| $Sr_{0.9}Ce_{0.1}Fe_{0.9}Ni_{0.1}O_3$ | SCeFN91 | | | | | 308 |
| $Sr_{0.8}Ce_{0.2}MnO_3$ | SCeM | | | | | 309,310 |
| $Sr_{0.8}Ce_{0.2}Mn_{0.8}Co_{0.2}O_3$ | SCeMC | | | | | 309,310 |
| $SrCo_{0.85}Fe_{0.15}O_3$ | SCF15 | | | | | 311 |
| $SrCo_{0.4}Fe_{0.5}Nb_{0.1}O_3$ | SCFN451 | | | | | 312 |
| $SrCo_{0.7}Fe_{0.2}Nb_{0.1}O_3$ | SCFN721 | | | | | 312,313 |
| $SrCo_{0.85}Fe_{0.1}P_{0.05}O_3$ | SCFP | | | | | 311 |
| $SrCo_{0.9}Nb_{0.1}O_3$ | SCN10 | 138,314 | 138,314 | | | 312,315–317 |
| $SrCo_{0.8}Nb_{0.2}O_3$ | SCN20 | 318 | 318 | | | 318 |
| $SrCo_{0.8}Nb_{0.1}Ta_{0.1}O_3$ | SCNTa | | | | | 124,298 |
| $SrCoO_3$ | SCO | 319 | 319 | | | 86,159,319 |
| $SrCo_{0.8}Sc_{0.2}O_3$ | SCSc82 | 319 | 319 | | | 319–321 |
| $SrCo_{0.9}Sc_{0.1}O_3$ | SCSc91 | 319 | 319 | | | 319 |
| $SrCo_{0.75}Sc_{0.125}Nb_{0.125}O_3$ | SCScN | | 322 | | | 322,323 |
| $SrSc_{0.175}Nb_{0.025}Co_{0.8}O_3$ | SCScN80 | | | | | 132 |
| $SrCo_{0.9}Ti_{0.1}O_3$ | SCT10 | | | | | 324 |
| $SrCo_{0.8}Ti_{0.2}O_3$ | SCT20 | | | | | 324 |
| $SrCo_{0.9}Ta_{0.1}O_3$ | SCTa10 | | | | | 325 |
| $SrCo_{0.8}Ta_{0.2}O_3$ | SCTa20 | 318 | 318 | | | 318 |
| $SrCo_{0.6}Ta_{0.4}O_3$ | SCTa40 | | | | | 325 |
| $SrCo_{0.9}Y_{0.1}O_3$ | SCY10 | | | | | 326 |
| $SrCo_{0.8}Y_{0.2}O_3$ | SCY20 | | | | | 326 |
| $SrFe_{0.75}Cr_{0.25}O_3$ | SFCr25 | | | | | 327 |
| $SrFe_{0.9}Hf_{0.1}O_3$ | SFH10 | 328 | 328 | | | 328 |
| $SrFe_{0.8}Mo_{0.2}O_3$ | SFMo20 | | | | | 327 |
| $SrFe_{0.9}Nb_{0.1}O_3$ | SFN10 | | | | | 329–332 |
| $SrFe_{0.8}Nb_{0.2}O_3$ | SFN20 | | | | | 329,333 |
| $SrFe_{0.8}Nb_{0.1}Ta_{0.1}O_3$ | SFNTa | | | | | 333,334 |
| $SrFeO_3$ | SFO | 226,328 | 226,328 | 335 | | 86,308,327,328,335–337 |
| $SrSc_{0.175}Nb_{0.025}Fe_{0.8}O_3$ | SFScN80 | | | | | 338 |
| $SrSc_{0.025}Nb_{0.075}Fe_{0.9}O_3$ | SFScN90 | | | | | 339 |
| $SrFe_{0.9}Si_{0.1}O_3$ | SFSi | | | | | 337 |
| $SrFe_{0.9}Ti_{0.1}O_3$ | SFT10 | | | | | 340,341 |
| $SrFe_{0.8}Ta_{0.2}O_3$ | SFTa20 | | | | | 333 |
| $SrFe_{0.9}Zr_{0.1}O_3$ | SFZ10 | | | | | 336 |
| $SrFe_{0.8}Zr_{0.2}O_3$ | SFZ20 | | | | | 336 |
| $Sr_{0.9}K_{0.1}FeO_3$ | SKF | | | | | 342 |
| $Sr_{0.9}K_{0.1}Fe_{0.1}Co_{0.9}O_3$ | SKFC19 | | | | | 342 |
| $Sr_{0.9}K_{0.1}Fe_{0.3}Co_{0.7}O_3$ | SKFC37 | | | | | 342 |
| $Sr_{0.9}K_{0.1}Fe_{0.5}Co_{0.5}O_3$ | SKFC55 | | | | | 342 |
| $Sr_{0.95}K_{0.05}Fe_{0.8}Nb_{0.1}Ta_{0.1}O_3$ | SKFNTa | | | | | 334 |
| $Sr_{0.95}Li_{0.05}Fe_{0.8}Nb_{0.1}Ta_{0.1}O_3$ | SLiFNTa | | | | | 334 |
| $SmCoO_3$ | SmCO | | | 343 | 343 | 159 |
| $Sr_{0.95}Na_{0.05}Fe_{0.8}Nb_{0.1}Ta_{0.1}O_3$ | SNaFNTa | | | | | 334 |
| $Sm_{0.2}Sr_{0.8}CoO_3$ | SSC28 | | | | | 344 |
| $Sr_{0.5}Sm_{0.5}CoO_3$ | SSC55 | 345 | 345 | 343 | 343 | 159,344,346–350 |
| $Sm_{0.8}Sr_{0.2}CoO_3$ | SSC82 | | | 343 | 343 | 344 |
| $Sm_{0.5}Sr_{0.5}Co_{0.8}Fe_{0.2}O_3$ | SSCF5582 | | | | | 126,349 |
| $Sm_{0.5}Sr_{0.5}Co_{0.9}Nb_{0.1}O_3$ | SSCN | | | | | 347 |
| $Sm_{0.3}Sr_{0.7}Nb_{0.08}Co_{0.92}O_3$ | SSCN10 | | | | | 351 |
| $SmSrCo_2O_5$ | SSCO | | | | | 302 |
| $SrSc_{0.075}Ta_{0.025}Fe_{0.9}O_3$ | SScTaF | | | | | 352 |
| $Sm_{0.5}Sr_{0.5}FeO_3$ | SSF | | | | | 348 |
| $Sm_{0.5}Sr_{0.5}MnO_3$ | SSM | | | | | 346 |
| $Sm_{0.5}Sr_{0.5}Fe_{0.8}Cu_{0.2}O_3$ | SSmFCu | | | | | 353 |
| $Sm_{0.5}Sr_{0.5}NiO_3$ | SSN | | | | | 348 |
| $SrTi_{0.95}Fe_{0.05}O_3$ | STF05 | | | 335 | | |
| $SrTi_{0.9}Fe_{0.1}O_3$ | STF10 | | | 335 | | |
| $SrTi_{0.65}Fe_{0.35}O_3$ | STF35 | | | 335 | | 335 |
| $SrTi_{0.5}Fe_{0.5}O_3$ | STF50 | | | 335 | | |
| $SrTi_{0.2}Fe_{0.8}O_3$ | STF80 | | | 335 | | 335 |
| $Sr_{0.7}Y_{0.3}CoO_3$ | SYC | | | | | 354 |
| $YBaCo_2O_5$ | YBCO | | | | | 127,355,356 |
| $YBaFe_2O_5$ | YBFO | | | | | 154 |
| $YBa_{0.5}Sr_{0.5}Co_2O_5$ | YBSCO | | | | | 355 |

**Acknowledgements:** This project was funded by the United States Department of Energy, National Energy Technology Laboratory, in part, through a site support contract. Neither the United States Government nor any agency thereof, nor any of their employees, nor the support contractor, nor any of their employees, makes any warranty, express or implied, or assumes any legal liability or responsibility for the accuracy, completeness, or usefulness of any information, apparatus, product, or process disclosed, or represents that its use would not infringe privately owned rights. Reference herein to any specific commercial product, process, or service by trade name, trademark, manufacturer, or otherwise does not necessarily constitute or imply its endorsement, recommendation, or favoring by the United States Government or any agency thereof. The views and opinions of authors expressed herein do not necessarily state or reflect those of the United States Government or any agency thereof.

## Conflicts of Interest

The authors of have no conflicts of interest to declare.

## Data Availability

The database of collected experimental measurements and all data used to construct the figures is publicly available on Figshare: https://doi.org/10.6084/m9.figshare.24438079.v1.

# Supplementary Information for

**A Critical Assessment of Electronic Structure Descriptors for Predicting Perovskite Catalytic Properties**

**Authors:** Ryan Jacobs[1,*], Jian Liu[2,*], Harry Abernathy[2], Dane Morgan[1]

[1] Department of Materials Science and Engineering, University of Wisconsin-Madison, Madison, WI, 53706, USA.

[2] National Energy Technology Lab, Morgantown, WV, 26505, USA.

## Section S1: Additional details of experimental property database development

In this section, we provide additional discussion of an overall survey of basic statistics of each catalytic property in the database, analysis of flagging potentially problematic measurements, and enumerating the best-performing materials at various temperatures based on each catalytic property (e.g., highest $k_{chem}$ materials, lowest ASR materials). **Table *1*** in the main text provides a high-level summary of the data collected in this work.

Assembling a large database of perovskite catalytic properties provides us with useful information, such as knowledge of typical ranges of each property at a given temperature, the best-performing materials at various working temperatures, the ability to flag outlier data points potentially indicative of problematic measurements, and information related to typical experimental measurement uncertainties by averaging measurements of the same material from multiple studies. Before performing additional analysis, we first average the measurements for the cases where the same material was measured across multiple studies. This averaging also provides an error bar for the measurements of these materials, which we calculate as the standard deviation of the measurements. From this, we find that the average experimental error bars across all properties at 500 °C, 600 °C, 700 °C and 800 °C are 0.68, 0.54, 0.47 and 0.45 log units, respectively. The above temperature values were chosen as they represent a typical operational range of SOFC, SOEC, r-SOFC and PCFC devices, where 500 °C is a desirable low temperature target for next-generation fuel cell and electrolyzer materials, and 800 °C is a more conventional intermediate-to-high temperature historically assessed in many experiments.

**Table S3** below provides a summary of some basic statistics of each perovskite catalytic property investigated in this work, including the minimum value, maximum value, average value, and standard deviation of all acquired values for 500 °C, 600 °C, 700 °C and 800 °C. The corresponding histogram distributions of property values at 500 °C and 800 °C are provided in **Figure 2** of the main text. The information shown in **Table S3** and **Figure 2** of the main text enables observations of characteristics of the data which will be important for understanding assessments of trends in property values with the O p-band center as discussed in **Section 2.2** of the main text. First, the dynamic range, and therefore, the standard deviation, of the data is temperature-dependent. In general, the dynamic range and standard deviations of the data are larger at lower temperature for all five properties examined. For example, for the $k_{chem}$ data from ECR, at 500 °C and 800 °C the data range over about 5.5 and 3.5 log units, respectively, and have standard deviations of 1.15 and 0.84 log units, respectively. Second, the dynamic range of the k* and D* values are larger than the $k_{chem}$ and $D_{chem}$ values. For example, at 500 °C and 800 °C the k* values range over about 12 and 7 log units, respectively, which is about 3.5-5 log units larger than the $k_{chem}$ ranges quoted above. This larger dynamic range of k* data may make correlations with these data less susceptible to variations caused by experimental error (because the experimental error is a smaller fraction of the range of the data), which, as discussed above, can be on the order of 0.68 and 0.45 log units at 500 °C and 800 °C, respectively.

The property value distributions shown in **Figure 2** of the main text suggest that some measurements can be flagged as outliers as they may be potentially problematic. Here, we define a data point as being an outlier if its value is more than two standard deviations above or below the mean value of that property at a given temperature. For example, at 500 °C the minimum log k* value of -16.95 cm/s is more than two standard deviations below the mean value (-8.52 cm/s-2×2.16 cm/s = -12.84 cm/s). Based on this definition of outlier points, at 500 °C there are 5 (BFZ90, LSC82, LSCN8282, LSF64, LSFG82), 3 (LMO, LSF30, LSM15), 2 (LCaCr12 and LCaCr22), 1 (LCaCr35) and 13 (LCaCr22, LCrO, LFN82, LFO, LMO, LNF64, LSF20, LSM15, LSM20, NCO, PSFN7382, SmCO, SSM, SSN) outlier materials for $k_{chem}$, $D_{chem}$, k*, D*, and ASR, respectively. At 800 °C, there are 3 (LSCF6482, LSFG82, LSM35), 3 (LMO, LSM15, LSM20), 1 (LCaCr12), 0 and 13 (BCFY, CaMN91, LCaCr22, LCrO, LMO, LNF64, LSM15, LSM20, LSM35, LSMMg10, SmCO, SSM, SSN) outlier materials for $k_{chem}$, $D_{chem}$, k*, D*, and ASR, respectively. Note that all of the data are used to fits throughout the main text and **SI**, and the above statements regarding outlier points is meant to illustrate a useful application of the database in understanding properties of the data.

**Table S3.** Overview of database statistics for each perovskite catalytic property investigated for temperatures in range of 500-800 °C. The order of the values given are minimum, maximum, average and standard deviation. All values are listed as base-10 logarithm.

| | $k_{chem}$ (cm/s) | $D_{chem}$ (cm$^2$/s) | k* (cm/s) | D* (cm$^2$/s) | ASR (Ohm-cm$^2$) |
|---|---|---|---|---|---|
| **500 °C** | -7.68, -2.23, -4.64, 1.15 | -9.63, -3.75, -5.99, 1.36 | -16.95, -5.09, -8.51, 2.17 | -21.51, -6.33, -12.70, 4.37 | -1.22, 5.60, 0.93, 1.05 |
| **600 °C** | -6.55, -2.17, -3.87, 0.98 | -8.40, -3.27, -5.26, 1.16 | -14.36, -4.99, -7.60, 1.80 | -18.50, -5.67, -11.35, 3.79 | -1.52, 4.36, 0.02, 0.94 |
| **700 °C** | -5.65, -1.93, -3.25, 0.89 | -7.42, -2.89, -4.67, 1.04 | -12.29, -4.18, -6.87, 1.57 | -16.29, -4.79, -10.27, 3.35 | -2.44, 3.36, -0.69, 0.88 |
| **800 °C** | -4.92, -1.47, -2.75, 0.84 | -6.62, -2.59, -4.19, 0.96 | -10.61, -3.38, -6.28, 1.45 | -14.81, -4.07, -9.39, 3.03 | -3.19, 2.56, -1.28, 0.85 |

**Table S4** contains a summary of the top three best-performing materials in our database for each property for temperatures of 500 °C, 600 °C, 700 °C and 800 °C. We first discuss the top candidates with regard to $k_{chem}$ and $D_{chem}$ values. Based on these ECR measurements, $PrBaFe_2O_{5+\delta}$ (PBFO), $Bi_{0.7}Pr_{0.1}Ba_{0.2}FeO_3$ (BiPBF), $BaCo_{0.7}Fe_{0.22}Y_{0.08}O_3$ (BCFY), $BaFe_{0.4}Co_{0.4}Zr_{0.1}Y_{0.1}O_3$ (BFCZY), $BaCo_{0.6}Fe_{0.2}Sn_{0.1}Y_{0.1}O_3$ (BCFSnY) and $BaFe_{0.1}Zr_{0.9}O_3$ (BFZ90) are all top-performing materials. It is interesting that four of these six top materials are fully Ba A-site materials with a mixture of late transition metals and redox-inactive species like Zr, Y and Sn on the B-site. It is worth noting that stability and $k_{chem}$ trends investigated by Jacobs et al. for the BFCZ system with varying Zr content also support this observation.[63] Further, it is worth noting the high performance of BiPBF as a Co-free material, where eliminating Co is attractive both from the standpoint of cost and for its relative scarcity compared to other transition metals like Fe and Mn. Finally, while BFZ90 is the highest $k_{chem}$ material at low temperature, based on the outlier definition established above, it is anomalously high, indicating this material may benefit from additional study to confirm its potentially exceptional $k_{chem}$ value.

Next, regarding the top candidate k* and D* materials, the materials $Ba_{0.5}Sr_{0.5}Co_{0.8}Fe_{0.2}O_3$ (BSCF), $La_{0.3}Sr_{0.7}CoO_3$ (LSC37), $PrBaCo_2O_{5+\delta}$ (PBCO) and $Sm_{0.5}Sr_{0.5}CoO_3$ (SSC55) are the most frequently occurring. These materials are all majority Co-doped on the B-site and have a mixture of alkaline earth and rare earth metals on the A-site. It is interesting to note the lack of overlap between the top candidates from the k*/D* list compared to the $k_{chem}$/$D_{chem}$ list. From examining our database, this lack of overlap appears to be due to none of the top materials from the $k_{chem}$ and $D_{chem}$ lists being measured with IEDP (thus, no k* and D* measurements were made), as opposed to some fundamental difference in the underlying property measurement.

Finally, regarding ASR, many of the best materials like BCFY, $BaCo_{0.7}Fe_{0.22}Sc_{0.08}O_3$ (BCFSc), $Ba_{0.5}Sr_{0.5}Co_{0.7}Fe_{0.2}Ni_{0.1}O_3$ (BSCFN) are similar to the best $k_{chem}$/$D_{chem}$ materials, suggesting that high $k_{chem}$ and $D_{chem}$ values may be more important for realizing low ASR than high k* and D* values. In addition, the materials $Pr_{0.4}Sr_{0.6}Co_{0.2}Fe_{0.8}O_3$ (PSCF4628) and $Bi_{0.5}Sr_{0.5}Fe_{0.9}Sb_{0.1}O_3$ (BiSFSb10) differ from the other materials in that they do not contain majority Ba. These materials are also the best performing materials at 500 °C with an ASR already lower than the 0.1 Ohm-cm$^2$ threshold enabling the desirable performance target of 1 W/cm$^2$ cells (assuming a typical 1 A/cm$^2$ current). It is somewhat surprising that PSCF4628 performs so well at 500 °C, because it is compositionally very similar to $La_{0.6}Sr_{0.4}Co_{0.2}Fe_{0.8}O_3$ (LSCF6428), which has a much higher base-10 log ASR value of 1.23 at 500 °C. In addition, the very similar compound $Pr_{0.6}Sr_{0.4}Co_{0.2}Fe_{0.8}O_3$ (PSCF6428) has a base-10 log ASR value of 1.60 at 500 °C. It is worth noting that PSCF4628 has only a single measurement in our database, so obtaining additional data on this material to further verify its excellent performance may be warranted. Finally, the high performance of BiSFSb10 is noteworthy because, like BiPBF, it is a Co-free material with desirable catalytic properties.

**Table S4.** Summary of top three best-performing materials in the database for each property for temperatures in range of 500-800 °C. All values are listed as base-10 logarithm. Materials marked with an asterisk (*) were flagged as potentially problematic based on the outlier criteria of more than two standard deviations from the mean of the property data.

| | $k_{chem}$ (cm/s) | $D_{chem}$ (cm$^2$/s) | k* (cm/s) | D* (cm$^2$/s) | ASR (Ω-cm$^2$) |
|---|---|---|---|---|---|
| **500 °C** | *BFZ90: -2.23<br>BiPBF: -2.86<br>LSCrF2828: -3.07 | PBFO: -3.75<br>BiPBF: -3.81<br>BCFY: -4.42 | LSC37: -5.09<br>SSC55: -5.98<br>LSCN4664: -6.00 | LSC37: -6.33<br>BSCF: -6.79<br>PBCO: -7.84 | BiSFSb10: -1.22<br>PSCF4628: -1.04<br>BSCFN: -0.70 |
| **600 °C** | BFZ90: -2.17<br>BiPBF: -2.37<br>BSCN641: -2.48 | PBFO: -3.27<br>BiPBF: -3.36<br>BCFY: -3.76 | LSC37: -4.99<br>BSCF: -5.20<br>SSC55: -5.40 | BSCF: -5.67<br>LSC37: -6.19<br>PBCO: -7.13 | BCFY: -1.52<br>BSCFN: -1.45<br>BCFSc: -1.44 |
| **700 °C** | BSCN641: -1.93<br>BiPBF: -1.99<br>BFCZY: -2.08 | PBFO: -2.89<br>BiPBF: -3.01<br>BCFY: -3.23 | BSCF: -4.45<br>LSC37: -4.92<br>SSC55: -4.94 | BSCF: -4.79<br>LSC37: -6.08<br>PBCO: -6.56 | BCFY: -2.44<br>BCFSc: -2.17<br>SCeFN82: -2.06 |
| **800 °C** | BSCN641: -1.48<br>BFCZY: -1.56<br>BCFSnY: -1.59 | PBFO: -2.59<br>BiPBF: -2.72<br>BCFY: -2.80 | BSCF: -3.84<br>SSC55: -4.56<br>SSC82: -4.60 | BSCF: -4.08<br>LSC37: -5.98<br>PBCO: -6.10 | BCFY: -3.19<br>SCeFN82: -2.84<br>BCFSc: -2.76 |

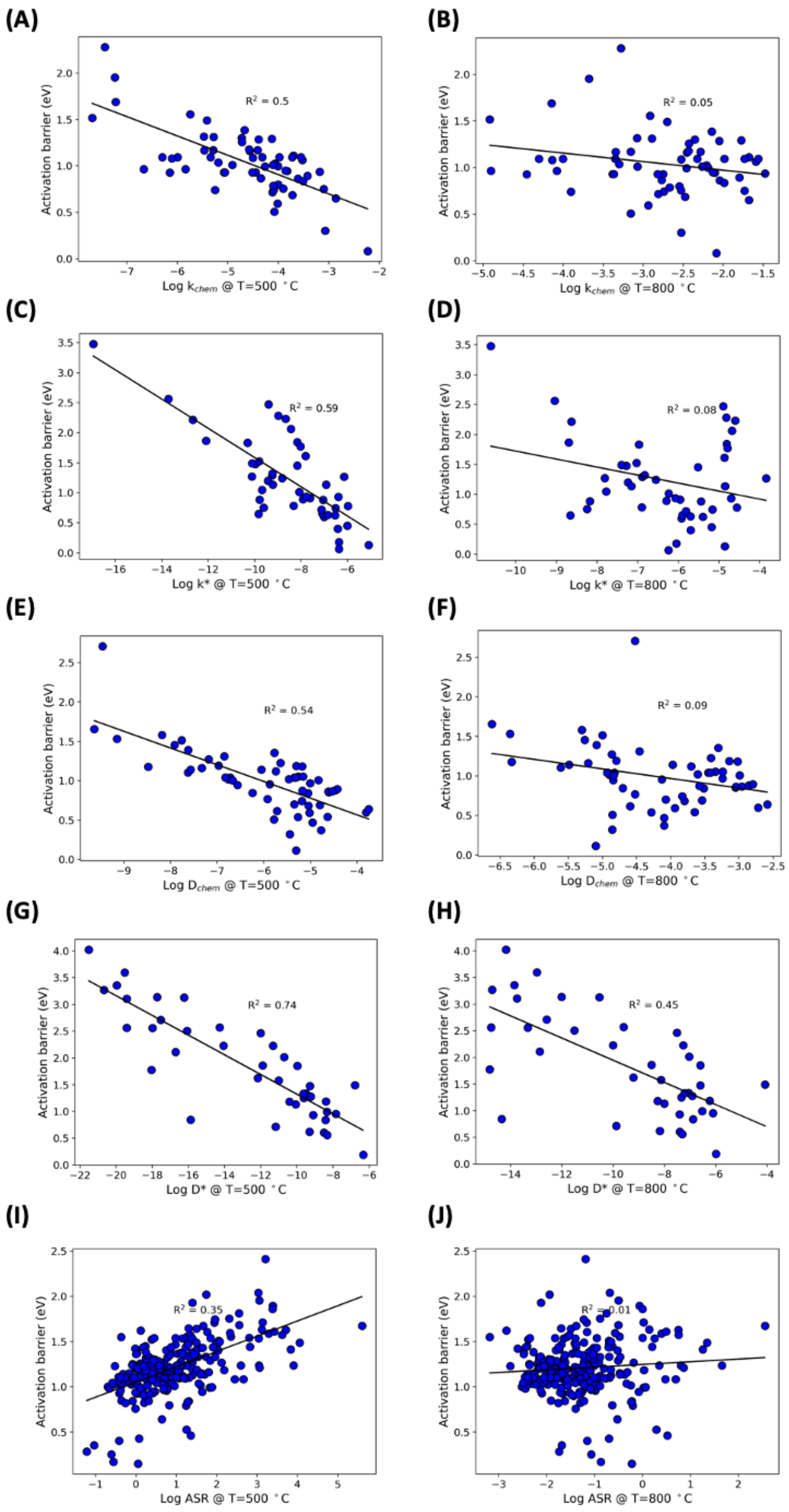

**Figure S7.** Plots of calculated activation energy barriers for each property, as a function of the property value at 500 °C and 800 °C. The left-hand column plots are all at 500 °C, while the right-hand column plots are at 800 °C. (A) and (B): $k_{chem}$, (C) and (D): k*, (E) and (F): $D_{chem}$, (G) and (H): D*, (I) and (J): ASR.

## Section S2: Additional details of O p-band center property trends

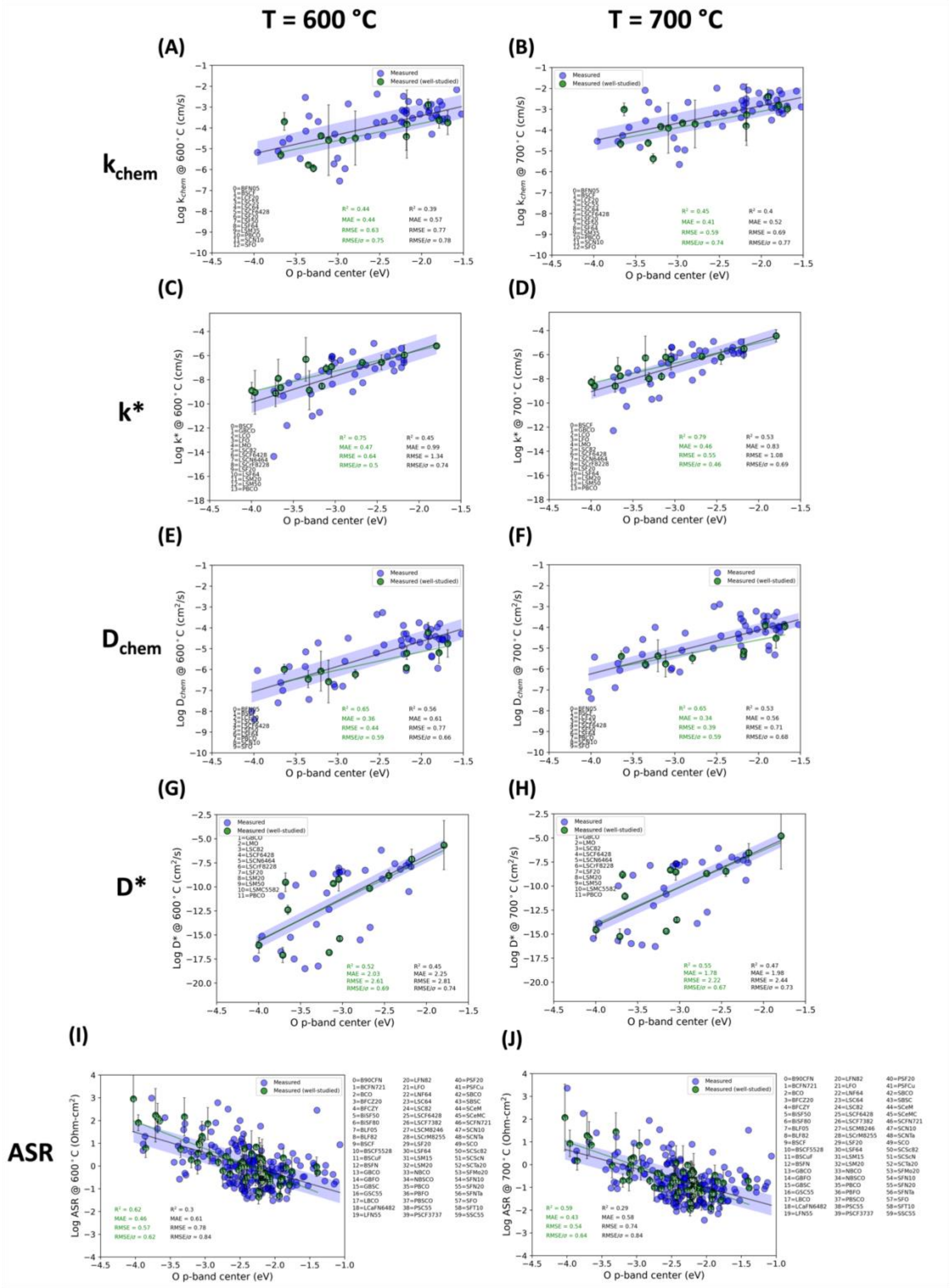

**Figure S8**. Trends of perovskite catalytic properties as a function of O p-band center. The left (right) column of plots is for T = 600 °C (T = 700 °C). The properties plotted are $k_{chem}$ (A, B), k* (C,

D), $D_{chem}$ (E, F), D* (G, H), and ASR (I, J). The blue points and black fit line and statistics are for a linear fit to all of the data. The green points and green fit line and statistics are for a linear fit to the subset of well-studied materials only (those materials with 2 or more measurements). The blue "x" points are outliers which were removed from the trend analysis. The blue shaded region is the average experimental error range, where the experimental error is calculated as the standard deviation of materials with multiple measurements. The average experimental error bars are 0.42 and 0.39 log units at 600 °C and 700 °C, respectively.

**Table S3.** Summary of O p-band linear fit slopes and intercepts.

| Property | Temperature (°C) | Data | Slope | Intercept |
|---|---|---|---|---|
| k* | 500 | All | 2.33 | -1.65 |
| | | Outliers removed | 1.74 | -3.14 |
| | | Well-studied only | 1.75 | -2.76 |
| | 600 | All | 2.20 | -1.12 |
| | | Outliers removed | 1.75 | -2.25 |
| | | Well-studied only | 1.71 | -2.17 |
| | 700 | All | 2.09 | -0.70 |
| | | Outliers removed | 1.76 | -1.55 |
| | | Well-studied only | 1.67 | -1.69 |
| | 800 | All | 2.01 | -0.36 |
| | | Outliers removed | 1.85 | -0.77 |
| | | Well-studied only | 1.64 | -1.30 |
| $k_{chem}$ | 500 | All | 1.01 | -2.12 |
| | | Outliers removed | 0.77 | -2.61 |
| | | Well-studied only | 0.87 | -2.91 |
| | 600 | All | 0.91 | -1.59 |
| | | Outliers removed | 0.80 | -1.78 |
| | | Well-studied only | 0.82 | -2.14 |
| | 700 | All | 0.83 | -1.17 |
| | | Outliers removed | 0.75 | -1.31 |
| | | Well-studied only | 0.78 | -1.53 |
| | 800 | All | 0.77 | -0.83 |
| | | Outliers removed | 0.66 | -1.02 |
| | | Well-studied only | 0.75 | -1.04 |
| D* | 500 | All | 4.98 | 2.40 |
| | | Outliers removed | 4.77 | 1.92 |
| | | Well-studied only | 4.73 | 1.35 |
| | 600 | All | 4.45 | 2.16 |
| | | Outliers removed | 4.45 | 2.16 |
| | | Well-studied only | 4.24 | 1.39 |
| | 700 | All | 4.04 | 1.98 |
| | | Outliers removed | 4.04 | 1.98 |
| | | Well-studied only | 3.84 | 1.41 |
| | 800 | All | 3.70 | 1.82 |
| | | Outliers removed | 3.70 | 1.82 |
| | | Well-studied only | 3.53 | 1.43 |
| $D_{chem}$ | 500 | All | 1.40 | -2.49 |
| | | Outliers removed | 1.15 | -3.00 |
| | | Well-studied only | 1.00 | -3.85 |
| | 600 | All | 1.20 | -2.24 |
| | | Outliers removed | 1.01 | -2.66 |
| | | Well-studied only | 0.87 | -3.41 |
| | 700 | All | 1.05 | -2.04 |
| | | Outliers removed | 0.86 | -2.44 |
| | | Well-studied only | 0.78 | -3.05 |
| | 800 | All | 0.92 | -1.89 |
| | | Outliers removed | 0.75 | -2.26 |
| | | Well-studied only | 0.70 | -2.77 |
| ASR | 500 | All | -1.00 | -1.45 |
| | | Outliers removed | -0.65 | -0.78 |
| | | Well-studied only | -1.36 | -2.30 |
| | 600 | All | -0.90 | -2.13 |
| | | Outliers removed | -0.65 | -1.63 |
| | | Well-studied only | -1.19 | -2.81 |
| | 700 | All | -0.83 | -2.68 |
| | | Outliers removed | -0.62 | -2.26 |
| | | Well-studied only | -1.05 | -3.22 |
| | 800 | All | -0.77 | -3.12 |
| | | Outliers removed | -0.55 | -2.66 |
| | | Well-studied only | -0.94 | -3.55 |

## Section S3: Additional details of catalytic property relationship trends

**Figure S9** contains plots assessing the above-mentioned correlations at both T = 500 °C (left column) and T = 800 °C (right column). All of these property correlations follow the same two general principles discussed at length in **Section 2.2** of the main text: the correlations are more robust at higher temperature, and more robust when considering fits to well-studied materials only. First, regarding the correlation of k* versus D* (**Figure S9A** and **Figure S9B**), the fit at 800 °C shows a very good linear correlation similar to the work of Kilner et al.,[65] but with a wider range of perovskite compositions here. At 500 °C, the correlation is modest and still quite good if one only considers well-studied materials.

Next, regarding the correlation of k* versus ASR (**Figure S9C** and **Figure S9D**), at both low (500 °C) and high (800 °C) temperatures the fits show reasonably good correlation, where k* can be predicted from ASR and vice versa with MAE values in the range of 0.5-0.7 log units, depending on the temperature and set of data considered. This trend is particularly useful as it directly relates k*, which is a materials property of the perovskite electrode catalyst, to ASR, which is related to the perovskite materials property but also includes additional confounding factors of the overall electrode device, such as its microstructure, the electrolyte it is bonded with, and any interfacial structure or phases at the electrode/electrolyte interface that may form.

Finally, we will refer to the relationship suggested by the Adler-Lane-Steele[66] formula, which is that $ASR=A\times(k^*\times D^*)^{-1/2}$ (A= constant) as the Adler-Lane-Steele relationship. The Adler-Lane-Steele formula was developed under very specific assumptions of mechanisms and microstructures, and it is not expected to apply to all the ASR measurements in this work. However, we explore it here just in case it provides a particularly robust relationship. Regarding the correlation of ASR in the Adler-Lane-Steele relationship (**Figure S9E** and **Figure S9F**), like the trend of k* versus ASR, there are reasonably good linear trends at both 500 °C and 800 °C. From **Figure S9E** and **Figure S9F**, when considering the subset of well-studied materials, the MAE values for predicting ASR from the Adler-Lane-Steele relationship are 0.55 and 0.37 log units at 500 °C and 800 °C, respectively. Inverting the correlations in **Figure S9C** and **Figure S9D** yields approximately the same MAE values of predicting ASR from just using k* of 0.58 and 0.28 log units at 500 °C and 800 °C, respectively (see **Figure S10**). One potential disadvantage of using the Adler-Lane-Steele relationship to predict new data is that it requires knowledge of both k* and D* to obtain a predicted ASR value, while the correlation of k* versus ASR does not require knowledge of D*.

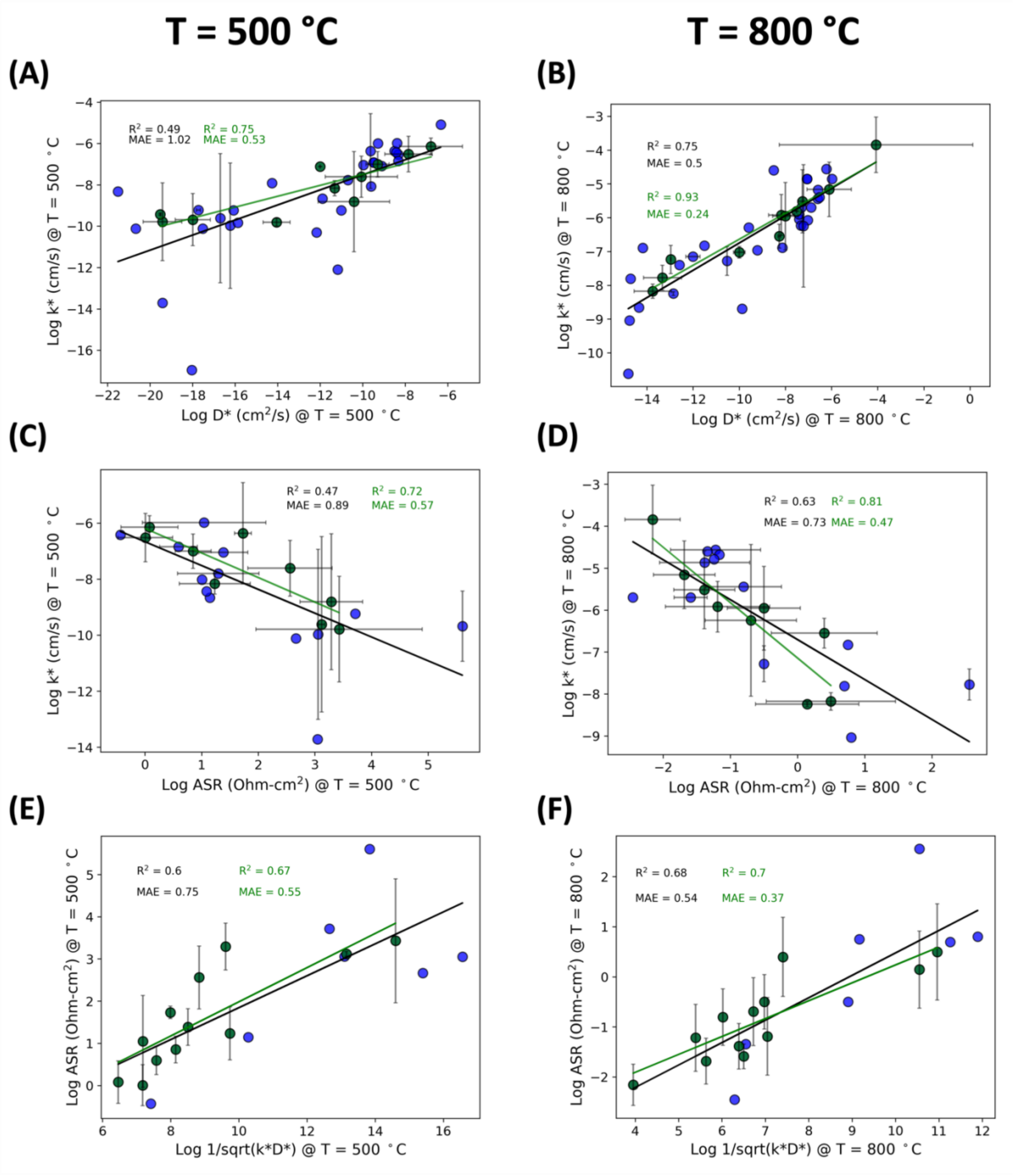

**Figure S9.** Assessments of correlations between perovskite catalytic properties. The left column of plots is at T = 500 °C, and the right column is at T = 800 °C. (A) and (B): Plots of k* vs. D* as proposed by Kilner et al.[65] at 500 °C and 800 °C, respectively. (C) and (D): Plots of k* vs. ASR at 500 °C and 800 °C, respectively, (E) and (F): Assessments of Adler-Lane-Steele[66] relationship at 500 °C and 800 °C, respectively.

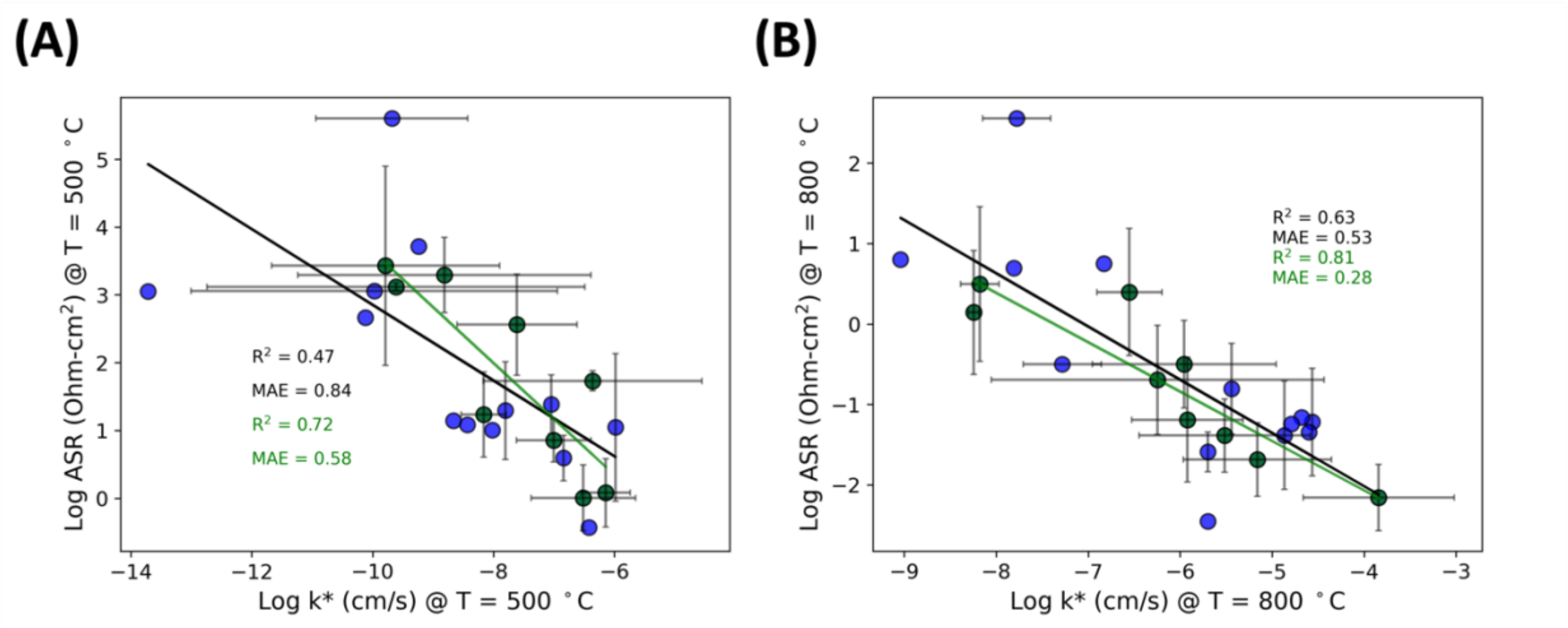

**Figure S10.** Plots of log k* vs. log ASR at 500 °C and 800 °C.

## Section S4: Additional thermodynamic factor analysis

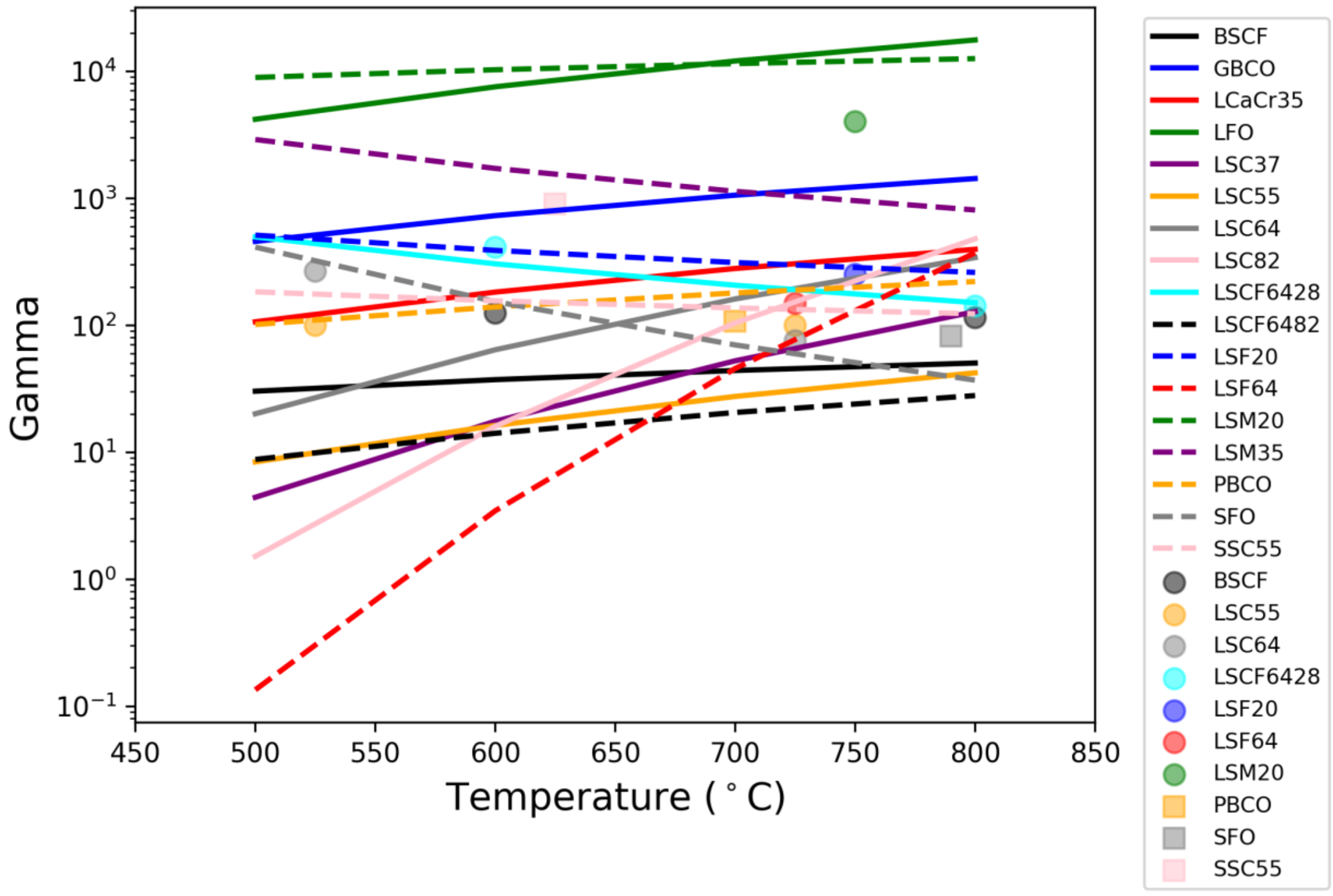

**Figure S5.** Plot of thermodynamic factor $\gamma$ as a function of temperature using extracted values of $k_{chem}/k^*$ at 500, 600, 700 and 800 °C. The data points signify experimentally measured $\gamma$ values at the designated temperature.